\documentclass[journal,10pt]{IEEEtran}
\ifCLASSINFOpdf
  \usepackage[pdftex]{graphicx}
  \usepackage{amsthm,amsmath,amssymb}
\usepackage{makecell}
\usepackage{mathrsfs}
  \usepackage{float}
  \usepackage{algorithm}
\usepackage{algpseudocode}
\usepackage{amsmath}
\usepackage{bm}
\usepackage{color}
\usepackage{comment}
\else
\fi
\usepackage{leftidx}
\usepackage{epstopdf}
\usepackage{amsmath}
\usepackage{subfigure}
\usepackage{makecell}
\usepackage{multirow}
\begin{document}

\title{A LiDAR-Aided Channel Model for Vehicular Intelligent Sensing-Communication  Integration}

\author{Ziwei~Huang,~\IEEEmembership{Graduate~Student~Member,~IEEE}, Lu~Bai,~\IEEEmembership{Member,~IEEE}, Mingran~Sun,~\IEEEmembership{Graduate~Student~Member,~IEEE},   and Xiang~Cheng,~\IEEEmembership{Fellow,~IEEE}

	\thanks{The authors would like to thank Boxun Liu and Yong Yu for their help in the collection of LiDAR point clouds in AirSim simulation platform.}

\thanks{Z.~Huang, M.~Sun, and X.~Cheng are with the State Key Laboratory of Advanced Optical Communication Systems and Networks, School of Electronics, Peking University, Beijing, 100871, P. R. China (email: ziweihuang@pku.edu.cn, mingransun@stu.pku.edu.cn, xiangcheng@pku.edu.cn).}
\thanks{L. Bai is with the Joint SDU-NTU Centre for Artificial Intelligence Research (C-FAIR), Shandong University, Jinan, 250101, P. R. China (e-mail: lubai@sdu.edu.cn).}
}

\markboth{}
{Zeng \MakeLowercase{\textit{et al.}}: Bare Demo of IEEEtran.cls for IEEE Journals}

		\maketitle

\begin{abstract}
In this paper, a novel channel modeling approach, named light detection and ranging (LiDAR)-aided  geometry-based stochastic modeling (LA-GBSM), is developed. Based on the developed LA-GBSM approach, 
a new millimeter wave (mmWave) channel model for sixth-generation (6G) vehicular intelligent sensing-communication integration
is proposed, which can support the design of  intelligent transportation systems (ITSs).
The proposed LA-GBSM is accurately parameterized under high, medium, and low vehicular traffic density (VTD) conditions via a sensing-communication simulation dataset with LiDAR point clouds and scatterer information  for the first time. Specifically, by detecting dynamic vehicles and static building/tress  through LiDAR point clouds via machine learning, scatterers  are divided into static and dynamic scatterers. Furthermore, statistical distributions of  parameters, e.g., distance, angle, number, and power, related to static and dynamic scatterers are quantified under high, medium, and low VTD conditions.  To mimic channel non-stationarity and consistency, based on the quantified statistical distributions, a new visibility region (VR)-based algorithm in consideration of newly generated static/dynamic scatterers is developed. Key channel statistics are derived and simulated. By comparing simulation results and ray-tracing (RT)-based results, the utility of the proposed LA-GBSM is verified.
\end{abstract}

\IEEEpeerreviewmaketitle

\section{Introduction}
\IEEEPARstart {I}{n} the upcoming sixth-generation (6G) era, the vehicular communication network, which is a key component in the intelligent transportation system (ITS), is expected to support more potential applications associated with future vehicles, including autonomous and smart vehicles. Bearing safety and efficiency as the utmost aims, autonomous and smart vehicles are equipped with  multi-modal sensors. These sensors are  wirelessly connected to provide comprehensive environmental feature  for the transportation purpose \cite{ISAC-11}.
%For example, by applying the millimeter wave (mmWave) technology with a  ultra-wide bandwidth, massive amounts of sensory information collected by multi-modal sensors can be exchanged among multiple vehicles in real time to achieve the beyond-vision-range sensing. 
Therefore, the integration of multi-modal sensing and communications is henceforth pivotal and natural in the vehicular communication network \cite{6G1}.
However, it cannot be adequately supported by the extensively studied integrated sensing and communications (ISAC), which limits to the integration of radio-frequency (RF) based radar sensing and communications.
To fill this gap, inspired by human
synesthesia, i.e., an involuntary human neuropsychological trait in which the stimulation of one human sense evokes another human sense, \cite{SOM} proposes a novel concept, i.e., \emph{Synesthesia of Machines (SoM)}. SoM refers to intelligent multi-modal sensing-communication  integration. Similar to how human senses the surrounding environment via multiple organs, multi-modal sensors and communication devices can also obtain environmental information, and thus we analogously refer to them as “machine senses” in SoM. Furthermore, the environmental information obtained by multiple organs can be mutually facilitated via complex biological neural networks in human synesthesia. Similarly,  in SoM,
the environmental information obtained by ``machine senses'' can be  processed via machine learning in an intelligent manner to achieve mutual facilitation.
 Unlike ISAC, which solely includes the integration of sensing and communications in RF, 
SoM refers to the intelligent integration of multi-modal sensing and communications, including 
 RF communications,  RF sensing, i.e., millimeter wave (mmWave) radar, and non-RF sensing, i.e., light detection and ranging (LiDAR) and RGB-D cameras etc. Therefore, SoM is a more generalized concept than ISAC and ISAC can be regarded as a special case of SoM. 
 As a typical application scenario in SoM, vehicular intelligent sensing-communication integration in ITSs has been received extensive attention. To support the design of vehicular intelligent sensing-communication integration systems,
 the proper characterization and modeling of 
  underlying channels  are essential \cite{SOM,myCOMST}.

Currently, the characterization and modeling of vehicular channels and the revealing of channel characteristics are based on vehicular channel measurement campaigns \cite{Heruisi}. The authors in \cite{Number2} carried out 
vehicle-to-vehicle (V2V) channel measurement campaigns at $5.2$ GHz in highway and rural scenarios in Lund, Sweden. In \cite{Number2},  time non-stationarity of rapidly-changing V2V channels was revealed, whereas the appearance and disappearance of scatterers were disregarded. The authors in \cite{Number3} conducted a different V2V channel measurement campaign at $5.9$ GHz in viaduct, tunnel, and cutting scenarios, where time non-stationarity of V2V channels and smooth scatterer  evolution in the time domain were explored. However, the existing V2V channel measurement campaigns, e.g., \cite{Number2} and \cite{Number3}, solely processed the RF  channel information collected by communication devices, without the facilitation of the sensory information collected by sensors. The lack of sensory information results in the fact that the transceiver cannot detect dynamic objects and static objects. In this case, it is significantly difficult to distinguish dynamic scatterers and static scatterers, which are certain objects, e.g., dynamic vehicles or static buildings/trees, in the physical environment \cite{liufan}--\cite{Zhang11}.
Due to the inability to  distinguish dynamic and static scatterers, the vehicular channel characteristics under different vehicular traffic density (VTD) conditions cannot be  explored. However, to support 6G applications for vehicular intelligent sensing-communication integration in ITSs, such as target localization and tracking for autonomous  vehicles, it is necessary to conduct precise environment modeling and explore vehicular channel characteristics under different VTD conditions \cite{6G1}.  Therefore, the existing vehicular channel measurement campaigns that solely process RF  channel information cannot adequately reveal channel characteristics.
%and support vehicular intelligent sensing-communication integration.

%In the physical environment, with the help of sensors, environmental features can be obtained, and static and dynamic objects can be detected and distinguished. In the propagation environment, it is essential to have an in-depth understanding of dynamic and static scatterers, which are certain objects, e.g., dynamic vehicles or static buildings/trees, in the physical environment \cite{liufan}--\cite{hengtai}. However, the existing V2V channel measurement campaigns, e.g., \cite{Number2} and \cite{Number3}, solely processed the  channel information collected by communication devices, without the facilitation of the sensory information collected by sensors. In such a condition, the understanding of static/dynamic scatterers generated by static/dynamic objects is limited. As a consequence, the vehicular channel characteristics under different vehicular traffic densities (VTDs), which significantly depend on static and dynamic scatterers and are  important physical features in vehicular  communications, cannot be explored in  the existing channel measurement campaigns.

Based on the vehicular channel characteristics revealed by channel measurement campaigns, extensive vehicular channel models have been proposed. To model time non-stationarity, the authors in \cite{NT1} and \cite{NT2} utilized the methods of capturing time-varying channel statistics and birth-death  process, respectively. 
However, the proposed models in \cite{NT1,NT2} cannot explore the impact of VTDs on  vehicular channel characteristics. 
To overcome this limitation, the authors in  \cite{yuan} proposed a V2V channel model and  attempted to explore the vehicular channel characteristics under high and low VTD conditions by dividing static  and dynamic scatterers. To further model time non-stationarity in consideration of VTDs, the authors in \cite{myTITS} proposed a V2V channel model and used the birth-death process to characterize the appearances and disappearances of static scatterers and dynamic scatterers, respectively.  When the mmWave technology is used  to high-mobility vehicular communications, 
 it is pivotal to further capture  time consistency and time-frequency non-stationarity \cite{myCOMST}. Towards this objective, our previous work in \cite{TWC_mixed} used VTD-based visibility region (VR)  and modeled frequency-dependent path gain for the components via static scatterers and dynamic scatterers. 
However, the existing vehicular channel models in \cite{yuan}--\cite{TWC_mixed} are solely at the stage of simply distinguishing static scatterers and dynamic scatterers for the modeling of VTDs. On the one hand, to model different VTD conditions, 
the existing vehicular channel models simply adjusted the numbers of dynamic and static scatterers, while ignored other key parameters, such as distance, angle, and delay-power. According to the standardized channel models, e.g., \cite{3GPP,5GCM}, the simultaneous investigation of number, distance, angle,  and delay-power parameters is important. On the other hand, since the determination of VR in different VTD conditions is not based on high-fidelity data, the smooth evolution of dynamic and static scatterers cannot be realistically characterized. Therefore, the existing vehicular channel models that mimic different VTD conditions by simply distinguishing dynamic and static scatterers cannot support the vehicular sensing and communication system design and autonomous driving. Specifically, various non-RF sensors, e.g., LiDAR, onboard autonomous vehicles can detect dynamic and static objects \cite{LiDAR}, and further align with the channel information collected by communication devices to distinguish dynamic and static scatterers under different VTD conditions. In such a condition, it is urgent to have a precise and comprehensive modeling of the parameters and smooth evolution of dynamic and static scatterers.

%However, vehicular intelligent sensing-communication integration  puts forward higher requirements for channel characterization and modeling. As a result, it is significantly necessary to conduct more in-depth exploration and modeling of VTDs and static/dynamic scatterers for channel characterization and modeling in vehicular intelligent sensing-communication integration. To be specific, environment modeling is the cornerstone of the research on vehicular intelligent sensing-communication integration in ITSs. Due to the introduction of sensors, the understanding of the physical environment is more comprehensive. In this case, a more accurate propagation environment modeling is also required. Specifically, sensors, such as light detection and ranging (LiDAR), have the ability to detect static and dynamic objects in the physical environment \cite{LiDAR}. These static and dynamic objects are candidates for static and dynamic scatterers in the propagation environment \cite{Zhang11}, which leads to the necessity of precisely modeling static/dynamic scatterers and VTDs.  Therefore, a realistic  channel model for vehicular intelligent sensing-communication integration in ITSs   is indispensable.
 
%An in-depth understanding and accurate characterization of the sensing-communication environment is the cornerstone of sensing-communication channel characterization and modeling. In this case,  it is necessary to  extract spatial features in physical environment and collect communication information in electromagnetic environment, which can be 

To fill this gap, we conduct a comprehensive channel characterization and modeling for vehicular intelligent sensing-communication integration based on a  sensing-communication dataset.
  Thanks to the intelligent integration of sensing and communications, sensory data  can be combined with communication data  to support channel characterization and modeling \cite{ISAC-11}.  Towards this objective,  a comprehensive sensing-communication  dataset, which contains sensory data and communication data, is essential. According to the dataset acquisition way, the existing datasets can be divided into measurement datasets and simulation datasets.  Measurement datasets are directly collected by the measurement equipment. Despite the measurement dataset, e.g., \cite{MEA}, can support the verification of fundamental methods, it is hard to flexibly customize the desired scenario because of the labor and cost
concerns. Unlike measurement datasets, the simulation dataset achieves a decent trade-off between complexity and fidelity via efficient software, and thus receives extensive attention.  Nonetheless, there is no software that can simultaneously collect sensory data and communication data. With the LiDAR simulator  in \cite{LIDARS} and Wireless InSite \cite{WI}, a sensing-communication simulation dataset, named LiDAR-COM in this paper, was constructed in \cite{new1}. In \cite{ViWi}, a different sensing-communication simulation dataset, named Vision-Wireless (ViWi), was proposed based on Wireless InSite and a popular game engine, i.e., Blender$^\mathrm{TM}$. However, the rendering of scenarios in the LiDAR-COM  dataset \cite{new1} and the ViWi dataset  \cite{ViWi} was relatively rough and the V2V channel data was ignored. To overcome these limitations, 
we constructed a  sensing-communication simulation dataset in \cite{dataset_cc} with superior rendering effects and V2V channel data.  The  sensing-communication dataset in \cite{dataset_cc} contains  high-fidelity sensory data collected by AirSim \cite{AirSim} and communication data collected by Wireless InSite \cite{WI}. 
With the help of the sensing-communication simulation dataset in \cite{dataset_cc},  inspired by the potential facilitation of sensory information on communications in SoM \cite{SOM},
a novel channel modeling approach, named LiDAR-aided geometry-based stochastic modeling (LA-GBSM),  is developed. The developed LA-GBSM  approach utilizes  LiDAR point clouds  to facilitate the revelation  of channel characteristics, and further leverages  geometry-based stochastic modeling (GBSM)  to model channels based on the revealed  channel characteristics. By using the developed LA-GBSM approach, 
a novel non-stationary and consistent mmWave channel model for vehicular intelligent sensing-communication integration is proposed. Although the proposed LA-GBSM utilizes single-modal sensory data, it  for the first time exploits non-RF sensory data, which is  processed by machine learning in an intelligent manner, 
to facilitate RF channel  modeling. Currently, the existing channel measurements and models solely process channel information and ignore the facilitation of sensory information, and thus cannot support SoM research.
The main contributions and novelties of this paper are summarized below.
\begin{enumerate}
\item With the help of LiDAR point clouds that are intelligently processed by machine learning,  a novel channel modeling approach, i.e., LA-GBSM, is developed. Based on the LA-GBSM approach,
a new non-stationary and consistent mmWave  channel model is proposed for vehicular intelligent
sensing-communication integration under high, medium, and low VTD conditions. The proposed LA-GBSM is adequately parameterized by the high-fidelity sensing-communication  dataset in \cite{dataset_cc}.
\item  In the proposed LA-GBSM, the number, distance, angle, and power-delay statistical distributions of dynamic and static scatterers under high, medium, and low VTD conditions are obtained via LiDAR point clouds 
for the first time.  This fills the gap that the existing channel measurement campaign without the facilitation
of  sensory information  cannot obtain these statistical distributions as 
it is exceedingly difficult to divide dynamic and static  scatterers solely by Doppler information in RF channels.
\item To capture time non-stationarity and consistency, based on the obtained statistical distributions tailored for high, medium, and low VTD conditions,  a new VR-based algorithm in consideration of  newly generated static/dynamic clusters is developed to mimic smooth cluster evolution. Furthermore, a frequency-dependent factor is  introduced to model frequency non-stationarity. 
\item Key channel statistical properties, including time-frequency correlation function (TF-CF) and Doppler power spectral density (DPSD), are derived and simulated. The close agreement between simulation results and ray-tracing (RT)-based results is achieved to verify the proposed LA-GBSM.
\end{enumerate}

The remainder of this paper is organized as follows. Section~II describes the  scenario of the sensing-communication simulation dataset and the intelligent processing of data. Section~III presents   channel characterization under high, medium, and low VTD conditions. In Section~IV, a novel non-stationary and consistent LA-GBSM for vehicular intelligent sensing-communication integration is proposed. Key channel statistical properties are given  in Section~V. Section~VI presents the simulation result, which is further
compared with the RT-based result. Finally,  Section~VII draws the conclusions.

\section{Simulation Testing for Vehicular Intelligent Sensing-Communication Integration}
Based on the sensing-communication  simulation dataset in  \cite{dataset_cc},  LiDAR point clouds and channel data are intelligently processed to determine static and dynamic scatterers. 

\subsection{Simulation Dataset in Vehicular Urban Crossroad with Different VTD Conditions}
\begin{figure*}[!t]
		\centering	\includegraphics[width=0.99\textwidth]{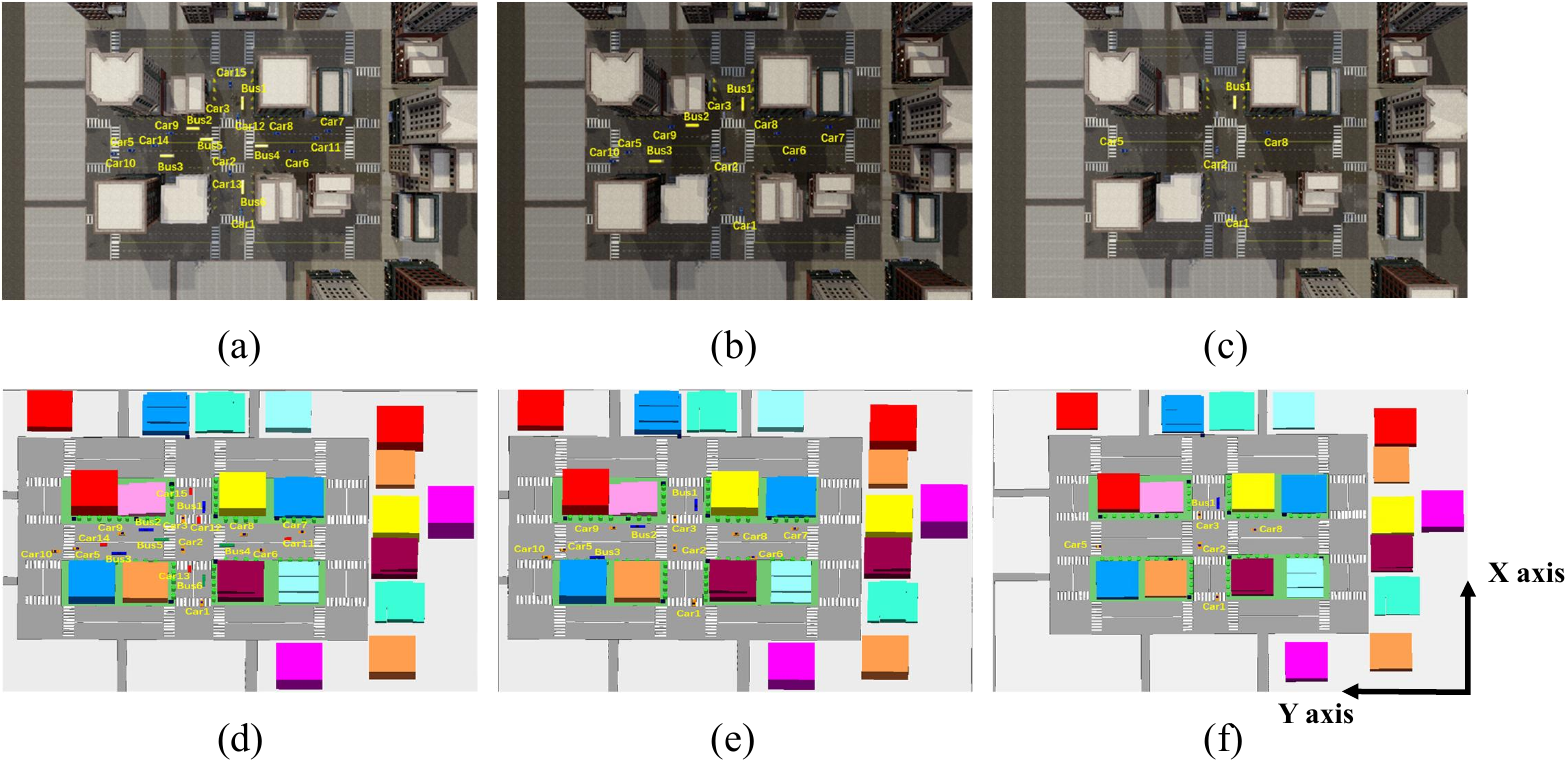}
	\caption{Vehicular simulation scenarios at the initial snapshot. Figs.~(a)--(c) are the scenarios in AirSim under high, medium, and low VTDs, respectively. Figs.~(d)--(f) are the scenarios in Wireless InSite under high, medium, and low VTDs, respectively.}
	\label{WI_AirSim}
	\end{figure*}

%As no software is tailored for constructing the  simulation dataset of sensing and communication integration, two different softwares, i.e., AirSim \cite{AirSim} and Wireless InSite \cite{WI}, are exploited to obtain sensory data in physical environment and communication data in electromagnetic environment, respectively. Based on the framework in \cite{dataset_cc}, the  in-depth integration of sensing and communications and the precise alignment between physical environment and electromagnetic environment are achieved. However, unlike \cite{dataset_cc}, 
In \cite{dataset_cc}, a high-fidelity sensing-communication simulation dataset was constructed based on two accurate simulation platforms, i.e., Wireless InSite \cite{WI} and AirSim \cite{AirSim}. 
Wireless InSite \cite{WI} uses RT technology based on geometrical optics and the uniform theory of diffraction to obtain accurate channel data. AirSim \cite{AirSim} constructs the simulation scenario  by using advanced three-dimensional (3D) modeling software with superior rendering effects, and thus can collect precise sensory data.
Here, a typical  urban crossroad scenario of vehicular intelligent sensing-communication integration in \cite{dataset_cc} with high, medium, and low VTD conditions is  considered.
The numbers of vehicles in high, medium, and low VTD conditions are 20, 12, and 8, respectively.
Fig.~\ref{WI_AirSim} depicts the  scenarios with high, medium, and low VTD conditions at the initial snapshot. The number of snapshots in each VTD condition is 300. Each car/bus is mobile
and is equipped with the LiDAR device and communication equipment.  The sensory data is collected by the LiDAR device, which has 16 channels, 10 Hz scanning frequency, 240,000 points per second, $-20^{\circ}\sim20^{\circ}$ horizontal field of view (FoV), and $-25^{\circ}\sim0^{\circ}$ vertical FoV. 
The communication data is obtained at a typical mmWave frequency band, i.e., $f_\mathrm{c}=28$~GHz carrier frequency with $2$~GHz  communication bandwidth. The numbers of antennas at transmitter (Tx) and receiver (Rx) are $M_\mathrm{T}=M_\mathrm{R}=1$. In each communication link, e.g., Car1 is the Tx and Car2 is the Rx, 
the 3D  coordinates of scatterers are obtained.
For clarity, Table~\ref{WI_link} gives the numbers of the obtained sensory and communication data.

\begin{table}[!t]
		\centering
		\caption{Numbers of LiDAR point clouds and links with scatterers under high, medium, and low VTD conditions.}
		\begin{tabular}{|c|c|c|}
			\hline
		\makecell[c]{\textbf{Condition}}	 &
	\makecell[c]{\textbf{LiDAR point clouds}} & 	\makecell[c]{\textbf{Links with scatterers}}	  	\\
			\hline
		High VTD	& 3,600 & 43,200 \\ 
	\hline
 		Medium VTD	& 3,000 & 30,000 \\ 
	\hline
  		Low VTD	& 1,800 & 10,800 \\ 
	\hline
   	\textbf{Total}	& \textbf{8,400} & \textbf{84,000} \\ 
	\hline
		\end{tabular}	
		\label{WI_link}
	\end{table}

\subsection{Intelligent Processing of Simulation Dataset }
In this subsection, steps of distinguishing dynamic  scatterers and static scatterers are given.

\begin{figure}[!t]
	\centering
	\subfigure[]{\includegraphics[width=0.48\textwidth]{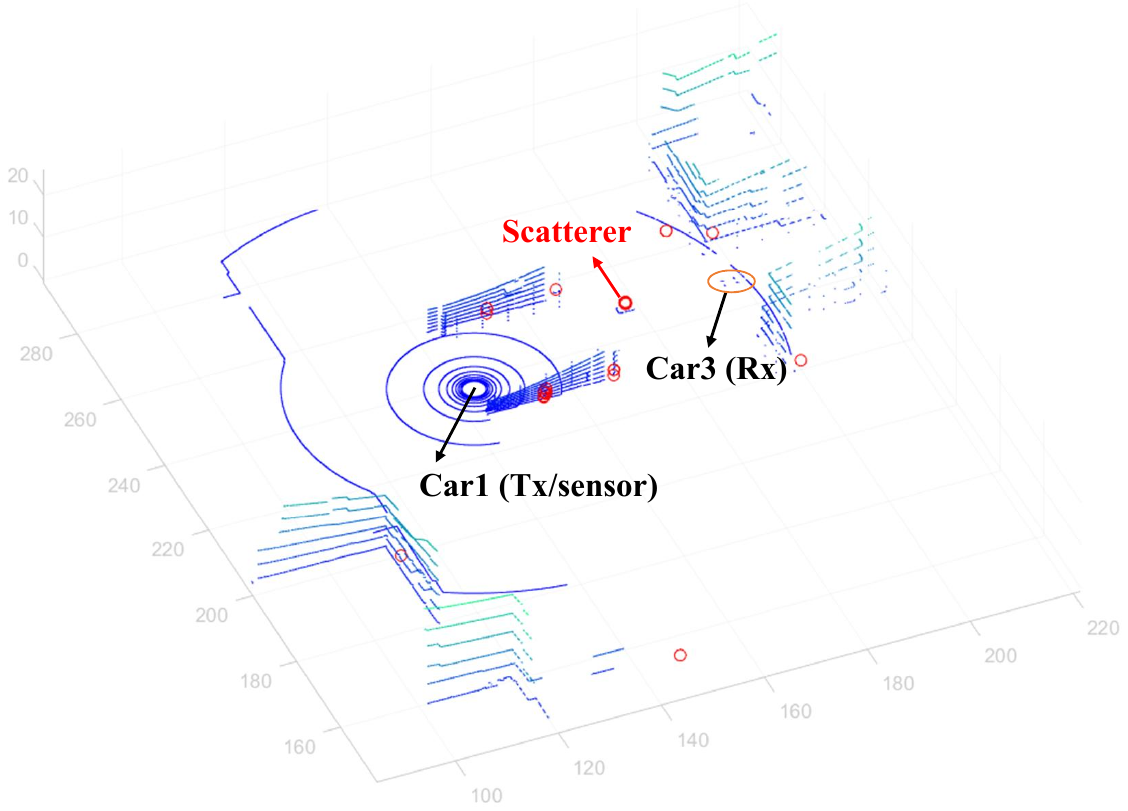}}
	\subfigure[]{\includegraphics[width=0.4\textwidth]{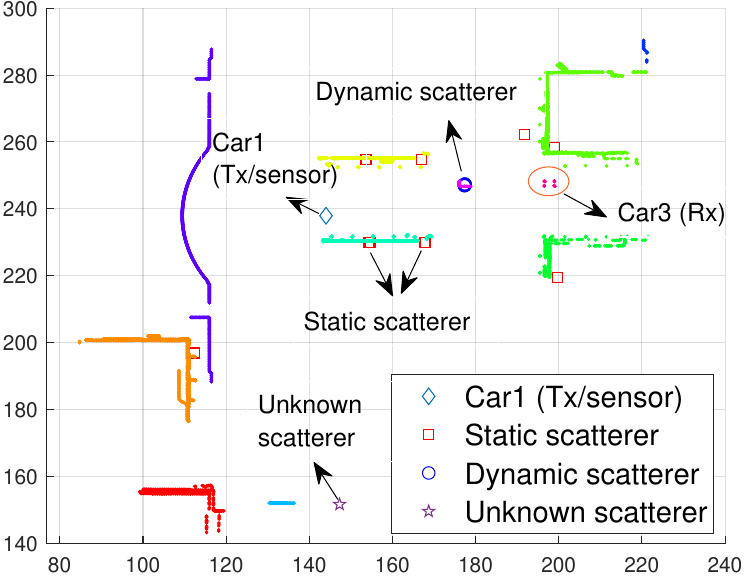}} 
	\caption{LiDAR point clouds and scatterers under medium VTD at initial time. (a)  LiDAR point clouds and scatterers before ground segmentation from 3D perspective. (b) LiDAR point clouds and scatterers after  DBSCAN clustering from the BEV perspective.}
 \label{match}
\end{figure}

The first step is to align electromagnetic environment with physical environment through coordinate transformation of LiDAR point clouds. The coordinate of sensor is given as $\mathbf{S}(t)=\left[x_{\mathrm {sen}}(t), y_{\mathrm {sen}}(t), z_{\mathrm {sen}}(t)\right]$ and the relative coordinate of the LiDAR point cloud is given as $\mathbf{L}'(t)=\left[x_{\mathrm {Li}}(t), y_{\mathrm{Li}}(t), z_{\mathrm{Li}}(t)\right]$. When the sensor moves in the positive $x$-axis, negative $x$-axis, positive $y$-axis, and 
negative $y$-axis directions, the absolute coordinates of the point cloud are respectively expressed as
\begin{equation}
\begin{footnotesize}
    \begin{aligned}
    \mathbf{L}_x^\mathrm{p}(t)=\left[ x_{\mathrm{Li}}(t)+x_{\mathrm {sen}}(t),  -y_{\mathrm{Li}}(t)-y_{\mathrm {sen}}(t),-z_{\mathrm{Li}}(t)-z_{\mathrm {sen}}(t)\right] 
      \end{aligned}
    \end{footnotesize}
\end{equation}
    \begin{equation}
\begin{footnotesize}
    \begin{aligned}
    \mathbf{L}_x^\mathrm{n}(t)=\left[ -x_{\mathrm{Li}}(t)+x_{\mathrm {sen}}(t),  y_{\mathrm{Li}}(t)+y_{\mathrm {sen}}(t),-z_{\mathrm{Li}}(t)-z_{\mathrm {sen}}(t)\right]
      \end{aligned}
    \end{footnotesize}
\end{equation}
    \begin{equation}
\begin{footnotesize}
    \begin{aligned}
         \mathbf{L}_y^\mathrm{p}(t)=\left[ y_{\mathrm{Li}}(t)+x_{\mathrm {sen}}(t), x_{\mathrm{Li}}(t)+y_{\mathrm {sen}}(t),-z_{\mathrm{Li}}(t)-z_{\mathrm {sen}}(t)\right] 
          \end{aligned}
    \end{footnotesize}
\end{equation}
    \begin{equation}
\begin{scriptsize}
    \begin{aligned}
         \mathbf{L}_y^\mathrm{n}(t)=
        \left[ -y_{\mathrm{Li}}(t)+x_{\mathrm {sen}}(t), -x_{\mathrm{Li}}(t)+y_{\mathrm {sen}}(t),-z_{\mathrm{Li}}(t)-z_{\mathrm {sen}}(t)\right].
    \end{aligned}
    \end{scriptsize}
\end{equation}
Fig.~\ref{match}(a) depicts LiDAR point clouds and scatterers under medium VTDs at  initial time, where Car3 is the Rx and Car1 is equipped with the LiDAR device and is also the Tx. It can be  seen from  Fig.~\ref{match}(a) that scatterers are generally located on LiDAR point clouds, and thus the alignment between electromagnetic environment and physical environment is achieved. 

The second step is the object classification based on the  coordinates of  LiDAR point clouds. However, extensive LiDAR points are ground points, which appear as a texture and significantly interfere with object point cloud
processing. To reduce the amount of data and improve the accuracy of the object classification, the ground point needs to be removed via the pre-processing of LiDAR point clouds. 
%Towards this objective, we use the elevation information of  LiDAR point clouds for ground segmentation \cite{Ground}. 
Then,  LiDAR point clouds solely include the information related to static objects
and dynamic vehicles in the environment.

The third step is to distinguish between dynamic vehicles and static objects based on  LiDAR point clouds. A typical \emph{machine learning algorithm}, i.e., DBSCAN clustering, is exploited. According to the clustering result, 
 the maximum and minimum values of the $x/y/z$-axis coordinates in a cluster are calculated. 
Since the size of vehicle is small, it is detected as a dynamic vehicle
if the difference between the maximum and minimum values of the $x$, $y$, and $z$-axis coordinates for a cluster is less than the size threshold on the $x$, $y$, and $z$-axis, respectively. Otherwise, the cluster is detected as a static object,  such as roadside trees and buildings.

The fourth step is to determine static scatterers  and dynamic scatterers in electromagnetic environment. Based on the detected dynamic/static objects and aligned electromagnetic  and physical environments, we classify scatterers as dynamic scatterers, static scatterers, and unknown scatterers. If the scatterer coincides with the detected  dynamic/static object, it is a dynamic/static scatterer. The remaining scatterers  are named unknown scatterers, which do not coincide with LiDAR points. The physical mechanism underlying the appearance of unknown scatterers is that they are beyond the detection range of the LiDAR device. In general, the unknown scatterer is  far away from the transceiver, and thus  can be ignored in channel realization due to the high path loss under mmWave communications. Fig.~\ref{match}(b) shows LiDAR point clouds and scatterers  after DBSCAN clustering from the 
 bird’s eye view (BEV) perspective, where  Car3 is the Rx and
Car1 is the Tx and is also equipped with the LiDAR device. In Fig.~\ref{match}(b),  vehicles are precisely detected and there are static, dynamic, and unknown scatterers. Therefore, in the fourth step, the 3D coordinates of static and dynamic scatterers can be obtained.

\section{Vehicular Intelligent Sensing-Communication Integration  Channel Characterization}
Currently, it is exceedingly difficult to distinguish  static and dynamic scatterers via channel measurement campaigns as they solely process Doppler information in RF channels, i.e.,  without the facilitation
of sensory information. In such a case, 
there is  no vehicular channel measurement campaign that  can   adequately characterize channels for  vehicular intelligent sensing-communication integration under different VTD conditions. Nevertheless, the accurate modeling of VTDs is of paramount importance for the proper channel characterization \cite{VTDsss}.
To fill this gap, by intelligently processing LiDAR point clouds and channel information, the statistical distributions of parameters related to static and dynamic scatterers are  explored under different VTD conditions. The statistical
values are summarized in Table~\ref{Parameter_1}. Consequently, we tailor static/dynamic scatterer parameters for the  modeling of vehicular intelligent sensing-communication integration channels under different VTD conditions.

			\begin{table*}[!t]
		\centering
		\caption{\textsc{Key statistical parameters in   V2V sensing and communication integration channels}}
		\label{Parameter_1}
				\begin{footnotesize}
		\renewcommand\arraystretch{1}
		\begin{tabular}{|c|c|c|c|c|c|}
			\hline
			\multicolumn{2}{|c|}{Parameter}	& Distribution & Type  & VTD & Value  \\
						\hline
      		\multicolumn{2}{|c|}{\multirow{12}{*}{Number}} & \multirow{12}{*}{Logistic}  & \multirow{3}{*}{Static cluster} & High & $\mu^\mathrm{c,L}_\mathrm{s}=0.09$, $\gamma^\mathrm{c,L}_\mathrm{s}=0.03$ \\
  \cline{5-6}
  \multicolumn{2}{|c|}{} & & &  Medium & $\mu^\mathrm{c,L}_\mathrm{s}=0.14$, $\gamma^\mathrm{c,L}_\mathrm{s}=0.04$\\
    \cline{5-6}
  \multicolumn{2}{|c|}{}& & &  Low & $\mu^\mathrm{c,L}_\mathrm{s}=0.08$, $\gamma^\mathrm{c,L}_\mathrm{s}=0.03$\\
    \cline{4-6}
  \multicolumn{2}{|c|}{}& & \multirow{3}{*}{Dynamic cluster} & High & $\mu^\mathrm{c,L}_\mathrm{d}=0.12$, $\gamma^\mathrm{c,L}_\mathrm{d}=0.06$\\
\cline{5-6}
\multicolumn{2}{|c|}{}& & & Medium & $\mu^\mathrm{c,L}_\mathrm{d}=0.09$, $\gamma^\mathrm{c,L}_\mathrm{d}=0.03$\\
    \cline{5-6}
\multicolumn{2}{|c|}{}& & & Low & $\mu^\mathrm{c,L}_\mathrm{d}=0.06$, $\gamma^\mathrm{c,L}_\mathrm{d}=0.03$\\
  \cline{4-6}
 \multicolumn{2}{|c|}{} & & \multirow{3}{*}{Static scatterer} & High & $\mu^\mathrm{s,L}_\mathrm{s}=0.45$, $\gamma^\mathrm{s,L}_\mathrm{s}=0.15$\\
  \cline{5-6}
\multicolumn{2}{|c|}{}  & & & Medium & $\mu^\mathrm{s,L}_\mathrm{s}=0.82$, $\gamma^\mathrm{s,L}_\mathrm{s}=0.23$\\
    \cline{5-6}
\multicolumn{2}{|c|}{}  & & & Low & $\mu^\mathrm{s,L}_\mathrm{s}=0.49$, $\gamma^\mathrm{s,L}_\mathrm{s}=0.16$\\
    \cline{4-6}
\multicolumn{2}{|c|}{}    & & \multirow{3}{*}{Dynamic scatterer} & High & $\mu^\mathrm{s,L}_\mathrm{d}=0.53$, $\gamma^\mathrm{s,L}_\mathrm{d}=0.25$\\
\cline{5-6}
\multicolumn{2}{|c|}{}& & & Medium & $\mu^\mathrm{s,L}_\mathrm{d}=0.44$, $\gamma^\mathrm{s,L}_\mathrm{d}=0.18$\\
    \cline{5-6}
\multicolumn{2}{|c|}{}& & & Low & $\mu^\mathrm{s,L}_\mathrm{d}=0.28$, $\gamma^\mathrm{s,L}_\mathrm{d}=0.17$\\
    \hline
\multicolumn{2}{|c|}{\multirow{6}{*}{Distance}}  & \multirow{3}{*}{Gamma}  & \multirow{3}{*}{Static scatterer} & High & $\alpha^\mathrm{G}_\mathrm{s}=0.68$, $\beta^\mathrm{G}_\mathrm{s}=1.74$ \\
  \cline{5-6}
\multicolumn{2}{|c|}{} & & &  Medium & $\alpha^\mathrm{G}_\mathrm{s}=0.83$, $\beta^\mathrm{G}_\mathrm{s}=1.71$\\
    \cline{5-6}
\multicolumn{2}{|c|}{}  & & &  Low & $\alpha^\mathrm{G}_\mathrm{s}=0.59$, $\beta^\mathrm{G}_\mathrm{s}=2.08$\\
    \cline{3-6}
\multicolumn{2}{|c|}{}& \multirow{3}{*}{Rayleigh}  & \multirow{3}{*}{Dynamic scatterer} & High & $\sigma^\mathrm{R}_\mathrm{d}=0.55$ \\
  \cline{5-6}
 \multicolumn{2}{|c|}{} & & &  Medium & $\sigma^\mathrm{R}_\mathrm{d}=0.37$\\
    \cline{5-6}
  \multicolumn{2}{|c|}{} & & &  Low & $\sigma^\mathrm{R}_\mathrm{d}=0.30$\\
    \hline  
      \multirow{24}{*}{Angle} & \multirow{6}{*}{AAoD} & \multirow{6}{*}{Gaussian}  & \multirow{3}{*}{Static scatterer} & High & $\mu^\mathrm{AAoD}_\mathrm{s}=-0.48$, $\sigma^\mathrm{AAoD}_\mathrm{s}=1.85$ \\
  \cline{5-6}
  & & & &  Medium & $\mu^\mathrm{AAoD}_\mathrm{s}=-0.12$, $\sigma^\mathrm{AAoD}_\mathrm{s}=2.08$\\
    \cline{5-6}
  & & & &  Low & $\mu^\mathrm{AAoD}_\mathrm{s}=0.26$, $\sigma^\mathrm{AAoD}_\mathrm{s}=1.76$\\
        \cline{4-6} 
       &  &   & \multirow{3}{*}{Dynamic scatterer} & High & $\mu^\mathrm{AAoD}_\mathrm{d}=-0.72$, $\sigma^\mathrm{AAoD}_\mathrm{d}=1.98$ \\
  \cline{5-6}
  & & & &  Medium & $\mu^\mathrm{AAoD}_\mathrm{d}=-0.54$, $\sigma^\mathrm{AAoD}_\mathrm{d}=1.78$\\
    \cline{5-6}
  & & & &  Low & $\mu^\mathrm{AAoD}_\mathrm{d}=-0.09$, $\sigma^\mathrm{AAoD}_\mathrm{d}=1.73$\\
    \cline{2-3}      \cline{4-6} 
     & \multirow{6}{*}{AAoA} & \multirow{6}{*}{Gaussian}  & \multirow{3}{*}{Static scatterer} & High & $\mu^\mathrm{AAoA}_\mathrm{s}=0.28$, $\sigma^\mathrm{AAoA}_\mathrm{s}=1.89$ \\
  \cline{5-6}
  & & & &  Medium & $\mu^\mathrm{AAoA}_\mathrm{s}=0.52$, $\sigma^\mathrm{AAoA}_\mathrm{s}=1.95$\\
    \cline{5-6}
  & & & &  Low & $\mu^\mathrm{AAoA}_\mathrm{s}=0.35$, $\sigma^\mathrm{AAoA}_\mathrm{s}=1.71$\\
       \cline{4-6}
     &  &  & \multirow{3}{*}{Dynamic scatterer} & High & $\mu^\mathrm{AAoA}_\mathrm{d}=0.62$, $\sigma^\mathrm{AAoA}_\mathrm{d}=1.98$ \\
  \cline{5-6}
  & & & &  Medium & $\mu^\mathrm{AAoA}_\mathrm{d}=0.81$, $\sigma^\mathrm{AAoA}_\mathrm{d}=1.61$\\
    \cline{5-6}
  & & & &  Low & $\mu^\mathrm{AAoA}_\mathrm{d}=1.01$, $\sigma^\mathrm{AAoA}_\mathrm{d}=1.58$\\
   \cline{2-3} \cline{4-6}
     & \multirow{6}{*}{EAoD} & \multirow{6}{*}{Gaussian} & \multirow{3}{*}{Static scatterer} & High & $\mu^\mathrm{EAoD}_\mathrm{s}=0.06$, $\sigma^\mathrm{EAoD}_\mathrm{s}=0.09$ \\
  \cline{5-6}
  & & & &  Medium & $\mu^\mathrm{EAoD}_\mathrm{s}=0.07$, $\sigma^\mathrm{EAoD}_\mathrm{s}=0.16$\\
    \cline{5-6}
  & & & &  Low & $\mu^\mathrm{EAoD}_\mathrm{s}=0.06$, $\sigma^\mathrm{EAoD}_\mathrm{s}=0.10$\\
       \cline{4-6}
     &  &  & \multirow{3}{*}{Dynamic scatterer} & High & $\mu^\mathrm{EAoD}_\mathrm{d}=0.57$, $\sigma^\mathrm{EAoD}_\mathrm{d}=0.62$ \\
  \cline{5-6}
  & & & &  Medium & $\mu^\mathrm{EAoD}_\mathrm{d}=0.66$, $\sigma^\mathrm{EAoD}_\mathrm{d}=0.59$\\
    \cline{5-6}
  & & & &  Low & $\mu^\mathrm{EAoD}_\mathrm{d}=0.80$, $\sigma^\mathrm{EAoD}_\mathrm{d}=0.56$\\
\cline{2-3} \cline{4-6}
     & \multirow{6}{*}{EAoA} &\multirow{6}{*}{Gaussian}  & \multirow{3}{*}{Static scatterer} & High & $\mu^\mathrm{EAoA}_\mathrm{s}=0.06$, $\sigma^\mathrm{EAoA}_\mathrm{s}=0.09$ \\
  \cline{5-6}
  & & & &  Medium & $\mu^\mathrm{EAoA}_\mathrm{s}=0.07$, $\sigma^\mathrm{EAoA}_\mathrm{s}=0.16$\\
    \cline{5-6}
  & & & &  Low & $\mu^\mathrm{EAoA}_\mathrm{s}=0.06$, $\sigma^\mathrm{EAoA}_\mathrm{s}=0.10$\\
       \cline{4-6}
     &  &  & \multirow{3}{*}{Dynamic scatterer} & High & $\mu^\mathrm{EAoA}_\mathrm{d}=0.57$, $\sigma^\mathrm{EAoA}_\mathrm{d}=0.62$ \\
  \cline{5-6}
  & & & &  Medium & $\mu^\mathrm{EAoA}_\mathrm{d}=0.66$, $\sigma^\mathrm{EAoA}_\mathrm{d}=0.59$\\
    \cline{5-6}
  & & & &  Low & $\mu^\mathrm{EAoA}_\mathrm{d}=0.80$, $\sigma^\mathrm{EAoA}_\mathrm{d}=0.56$\\
   \hline
\multicolumn{2}{|c|}{\multirow{6}{*}{Power-Delay}}  & \multirow{6}{*}{Exponential}  & \multirow{3}{*}{Static scatterer} & High & $\xi_\mathrm{s}=7.75\times10^6$, $\eta_\mathrm{s}=30.28$, $\sigma_\mathrm{E,s}=9.81$ \\
  \cline{5-6}
\multicolumn{2}{|c|}{} & & &  Medium & $\xi_\mathrm{s}=8\times10^6$, $\eta_\mathrm{s}=31.90$, $\sigma_\mathrm{E,s}=11.10$ \\
    \cline{5-6}
\multicolumn{2}{|c|}{}  & & &  Low & $\xi_\mathrm{s}=10\times10^6$, $\eta_\mathrm{s}=29.38$, $\sigma_\mathrm{E,s}=9.71$\\
    \cline{4-6}
\multicolumn{2}{|c|}{}&  & \multirow{3}{*}{Dynamic scatterer} & High & $\xi_\mathrm{d}=4.54\times10^6$, $\eta_\mathrm{d}=31.08$, $\sigma_\mathrm{E,d}=9.60$ \\
  \cline{5-6}
 \multicolumn{2}{|c|}{} & & &  Medium & $\xi_\mathrm{d}=1.50\times10^6$, $\eta_\mathrm{d}=32.80$, $\sigma_\mathrm{E,d}=10.90$ \\
    \cline{5-6}
  \multicolumn{2}{|c|}{} & & &  Low & $\xi_\mathrm{d}=4.47\times10^6$, $\eta_\mathrm{d}=30.17$, $\sigma_\mathrm{E,d}=8.72$ \\
    \hline  
		\end{tabular}
			\end{footnotesize}
	\end{table*}
 
\subsection{Scatterer and Cluster Numbers}

\begin{figure*}[!t]
	\centering
	\subfigure[]{\includegraphics[width=0.28\textwidth]{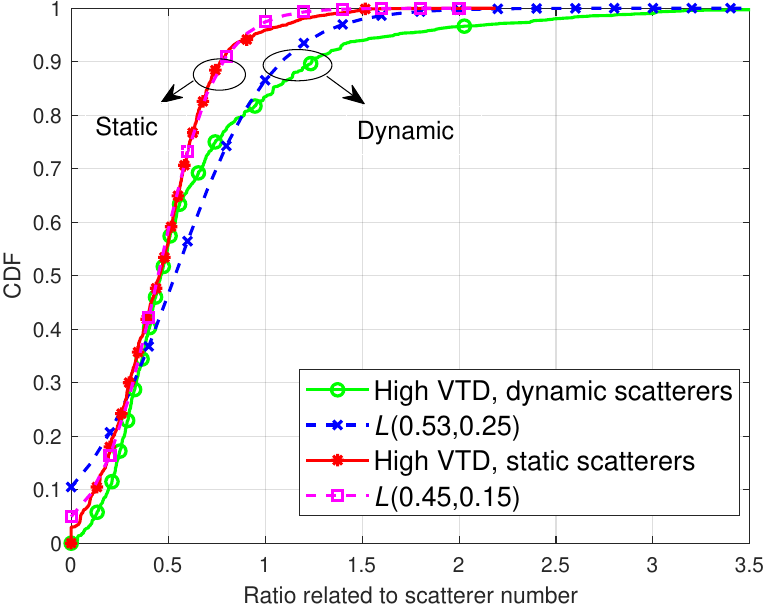}}
	\subfigure[]{\includegraphics[width=0.28\textwidth]{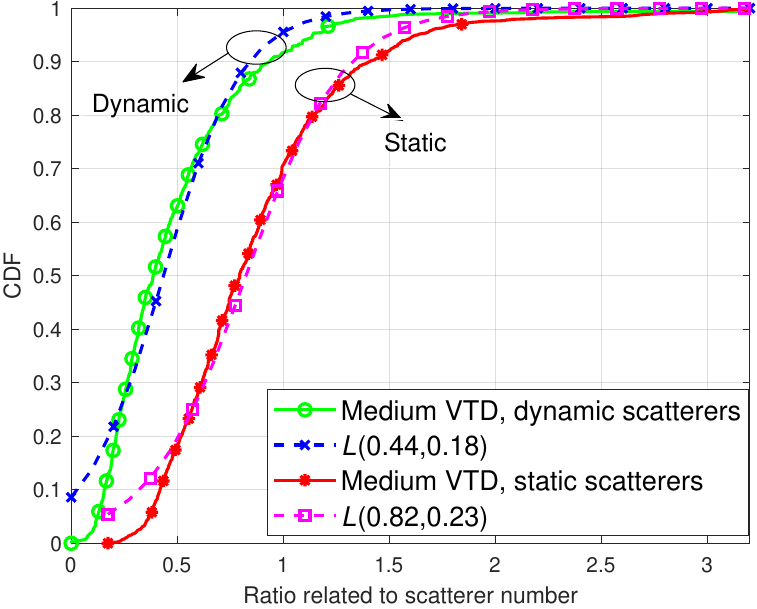}} 
   \subfigure[]{\includegraphics[width=0.28\textwidth]{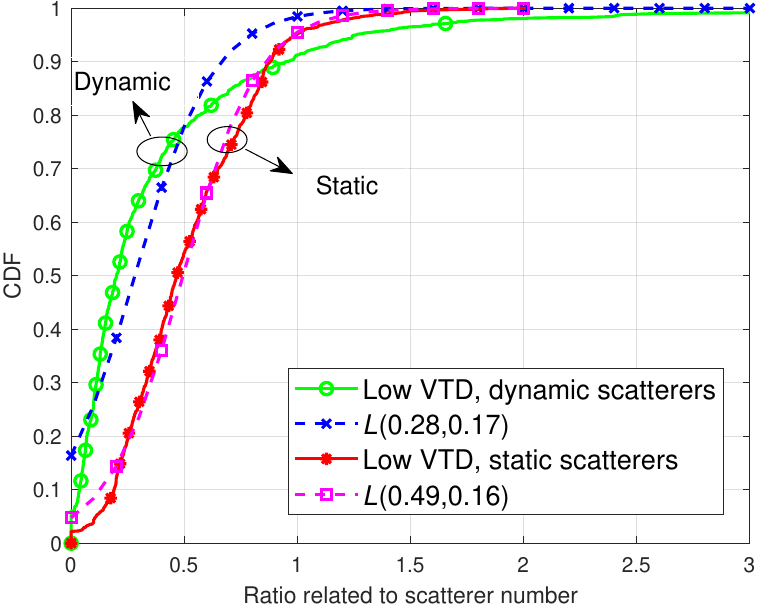}}
    \subfigure[]{\includegraphics[width=0.28\textwidth]{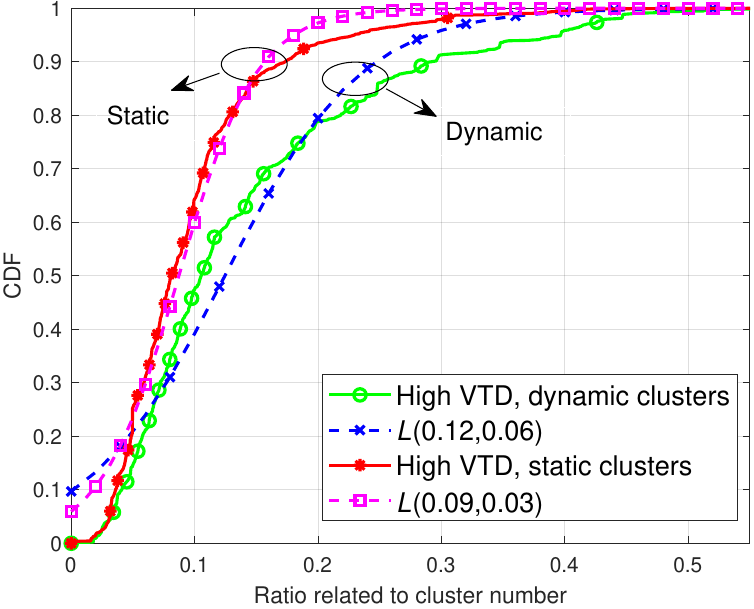}}  \subfigure[]{\includegraphics[width=0.28\textwidth]{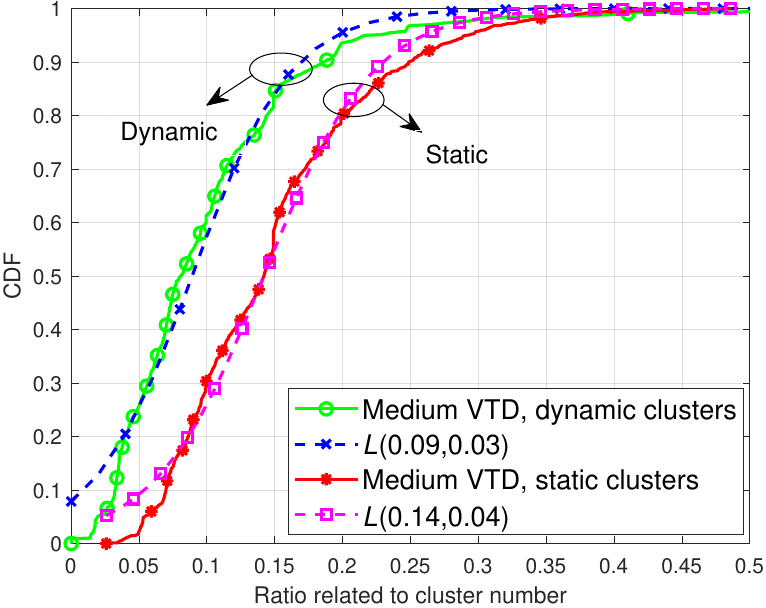}}
\subfigure[]{\includegraphics[width=0.28\textwidth]{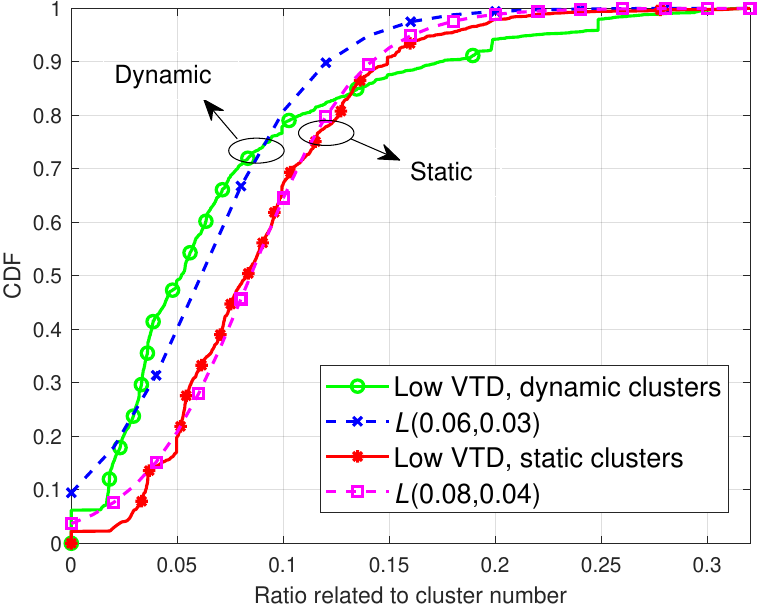}}
	\caption{CDFs of static/dynamic scatterer and cluster number parameters with the  Logistic  distribution fitting under different VTDs. Figs.~(a)--(c)  show the CDFs of scatterer number parameters under high, medium, and low VTDs, respectively. Figs.~(d)--(f)  show the CDFs of cluster number parameters under high, medium, and low VTDs, respectively.}
 \label{Number}
\end{figure*}

In  channel realization, the proper modeling of scatterer and cluster numbers is important \cite{Jiang11,hzw1}. However, in \cite{Number1},  the number of scatterers was modeled to obey the linear distribution and the distinction between dynamic scatterers and static scatterers was ignored.
Meanwhile, the authors in \cite{Number2} simply assumed that the numbers of static scatterers and dynamic scatterers obey the Gaussian distribution and the uniform distribution, respectively. To have a  comprehensive characterization of  channels for vehicular intelligent sensing-communication integration, 
 the numbers of static and dynamic scatterers are adequately explored. For generality, two new scatterer number parameters $N_\mathrm{s}^\mathrm{c1,c3}(t)$ and $M_\mathrm{s}^\mathrm{c1,c3}(t)$, which represent the   ratios of static and dynamic scatterer numbers to transceiver distance in a certain communication link, e.g., Car1 is the Tx and Car3 is the Rx, are introduced and written by
 \begin{equation}
     N_\mathrm{s}^\mathrm{c1,c3}(t)=\frac{I^\mathrm{c1,c3}(t)}{\|\mathbf{T}^\mathrm{c1,c3}(t)-\mathbf{R}^\mathrm{c1,c3}(t)\|}
 \end{equation}
  \begin{equation}
     M_\mathrm{s}^\mathrm{c1,c3}(t)=\frac{J^\mathrm{c1,c3}(t)}{\|\mathbf{T}^\mathrm{c1,c3}(t)-\mathbf{R}^\mathrm{c1,c3}(t)\|}
 \end{equation}
 where $I^\mathrm{c1,c3}(t)$ and  $J^\mathrm{c1,c3}(t)$ are the numbers of static and dynamic scatterers in the communication link between Car1, i.e., Tx, and Car3, i.e., Rx. $\mathbf{T}^\mathrm{c1,c3}(t)$ and $\mathbf{R}^\mathrm{c1,c3}(t)$ are  the positions of Car1 and Car3. Moreover, we compute  the number parameters of static  and dynamic scatterers for each communication link at each snapshot.
 Figs.~\ref{Number}(a)--(c) give the  cumulative distribution functions (CDFs)
 of all number parameters of static and dynamic scatterers  under high, medium, and low VTDs, respectively. It can be observed that the CDFs of static and dynamic scatterer number parameters can fit well with the Logistic  distribution. The CDF of the Logistic  distribution for static/dynamic scatterers is given by
 \begin{equation}
     F^\mathrm{s,L}_\mathrm{s/d}(x)=\frac{1}{1+e^{-(x-\mu^\mathrm{s,L}_\mathrm{s/d})/{\gamma^\mathrm{s,L}_\mathrm{s/d}}}}
 \end{equation}
 where $\mu^\mathrm{s,L}_\mathrm{s/d}$ is the mean and $\gamma^\mathrm{s,L}_\mathrm{s/d}$ is the scale parameter of the Logistic  distribution for static/dynamic scatterers.
In Table~\ref{Parameter_1} and Fig.~\ref{Number}, for the dynamic scatterer, the mean and variance of its number parameter increase as the VTD increases. This is because that the increase in the number of dynamic vehicles around the transceiver leads to the increase in the number of dynamic scatterers. 
Unlike dynamic scatterers, as the VTD increases, the mean and variance of static scatterer number parameter
increase first and then decrease. This phenomenon is explained that,
with the increase of vehicles, the environment tends to be a rich scattering environment (SE), i.e., there are more propagation paths. In this case, static components increase, leading to the increase in the number of  static scatterers. However, as the number of vehicles further increases, the static component is significantly blocked by vehicles, leading to the decrease in the number of  static scatterers. 

To further investigate channel characteristics for vehicular intelligent sensing-communication integration, the scatterers are clustered to explore the statistical distribution of  cluster number. Similarly, two new cluster number parameters $N_\mathrm{c}^\mathrm{c1,c3}(t)$ and $M_\mathrm{c}^\mathrm{c1,c3}(t)$, which represent the   ratios of static and dynamic clusters numbers to the distance between Car1, i.e., Tx, and Car3, i.e., Rx, are introduced. We calculate the number parameters of static  and dynamic
clusters for each communication link at each snapshot.
Figs.~\ref{Number}(d)--(f) present the CDFs of all number parameters of static and dynamic clusters
under high, medium, and low VTDs, respectively. The CDF of the Logistic  distribution for static/dynamic clusters can be written as
 \begin{equation}
     F^\mathrm{c,L}_\mathrm{s/d}(x)=\frac{1}{1+e^{-(x-\mu^\mathrm{c,L}_\mathrm{s/d})/{\gamma^\mathrm{c,L}_\mathrm{s/d}}}}
 \end{equation}
 where $\mu^\mathrm{c,L}_\mathrm{s/d}$ is the mean and $\gamma^\mathrm{c,L}_\mathrm{s/d}$ is the scale parameter of the Logistic  distribution for static/dynamic clusters.
From Table~\ref{Parameter_1} 
 and Fig.~\ref{Number}, it can be seen that the observation of static/dynamic cluster number parameters is similar to that of  static/dynamic scatterer number parameters.

\subsection{Distance Parameters}
\begin{figure*}[!t]
	\centering
	\subfigure[]{\includegraphics[width=0.28\textwidth]{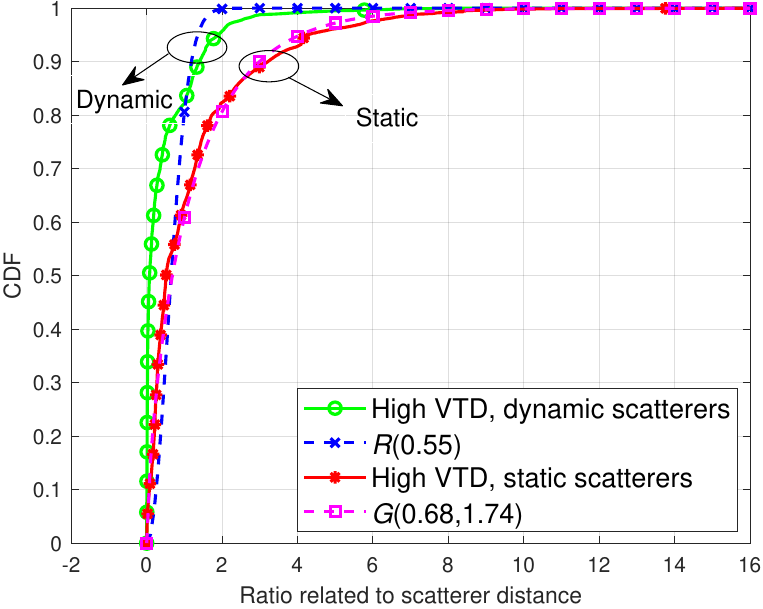}}
	\subfigure[]{\includegraphics[width=0.28\textwidth]{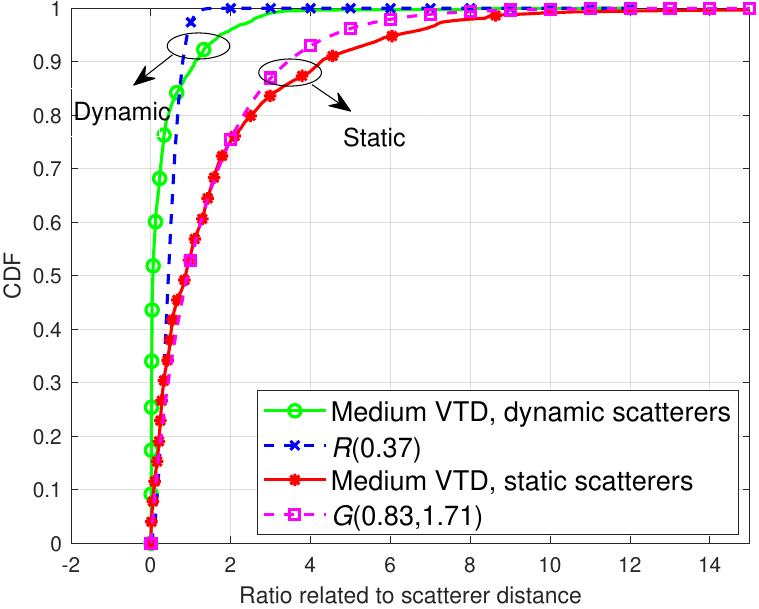}} 
   \subfigure[]{\includegraphics[width=0.28\textwidth]{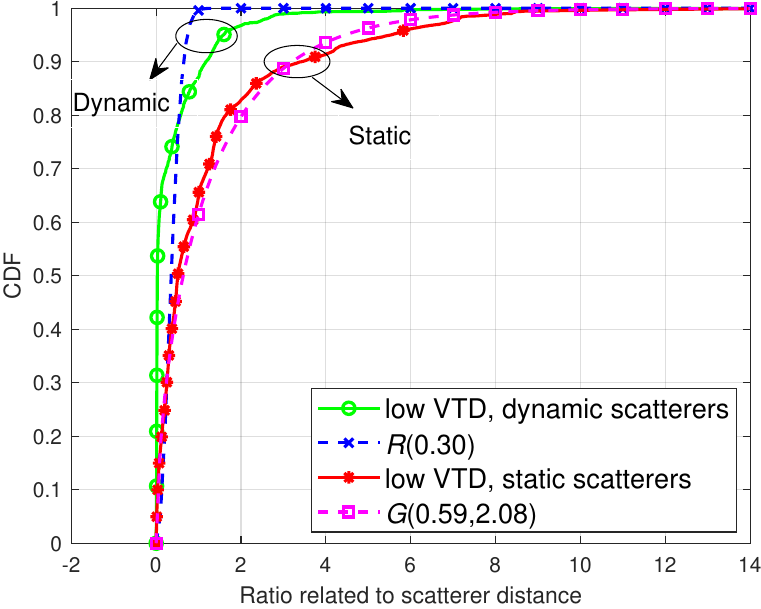}}
	\caption{CDFs of static/dynamic scatterer distance parameters with the Gamma/Rayleigh distribution fitting under different VTDs. (a) High VTD. (b) Medium VTD. (c) Low VTD.}
 \label{Distance}
\end{figure*}

In the stochastic channel modeling approach, distance parameters of scatterers can be assumed to obey a certain statistical distribution. The standardized channel models in \cite{3GPP} and \cite{Wu} assumed that distance parameters of scatterers obey the  Exponential distribution, while ignored the distinction between static and dynamic scatterers. To overcome this limitation, we simultaneously explore distance characteristics of static scatterers and dynamic scatterers. To be specific, two new distance parameters ${D}^\mathrm{c1,c3}_{i}(t)$ and ${D}^\mathrm{c1,c3}_{j}(t)$ related to the $i$-th static scatterer and the $j$-th dynamic scatterer are introduced and expressed as
\begin{equation}
\begin{tiny}
\begin{aligned}
 & {D}^\mathrm{c1,c3}_{i}(t)=\\
  &\frac{\left\|\mathbf{T}^\mathrm{c1,c3}(t)-\mathbf{S}_i^\mathrm{c1,c3}(t)\right\|+\left\|\mathbf{R}^\mathrm{c1,c3}(t)-\mathbf{S}_i^\mathrm{c1,c3}(t)\right\|-\|\mathbf{T}^\mathrm{c1,c3}(t)-\mathbf{R}^\mathrm{c1,c3}(t)\|}{\|\mathbf{T}^\mathrm{c1,c3}(t)-\mathbf{R}^\mathrm{c1,c3}(t)\|}
  \end{aligned}
  \end{tiny}
\end{equation}
\begin{equation}
\begin{tiny}
\begin{aligned}
&  {D}^\mathrm{c1,c3}_{j}(t)=  \\
&\frac{\left\|\mathbf{T}^\mathrm{c1,c3}(t)-\mathbf{S}_j^\mathrm{c1,c3}(t)\right\|+\left\|\mathbf{R}^\mathrm{c1,c3}(t)-\mathbf{S}_j^\mathrm{c1,c3}(t)\right\|-\|\mathbf{T}^\mathrm{c1,c3}(t)-\mathbf{R}^\mathrm{c1,c3}(t)\|}{\|\mathbf{T}^\mathrm{c1,c3}(t)-\mathbf{R}^\mathrm{c1,c3}(t)\|}
    \end{aligned}
  \end{tiny}
\end{equation}
where $\|\cdot\|$ denotes the Frobenius norm. $\mathbf{S}_i^\mathrm{c1,c3}(t)$ and  $\mathbf{S}_j^\mathrm{c1,c3}(t)$ denote the positions of the $i$-th static scatterer and the $j$-th dynamic scatterer in the communication link between Car1, i.e., Tx, and Car3, i.e., Rx. Furthermore, we compute  the distance parameter of each static/dynamic scatterer for each
communication link at each snapshot.
Figs.~\ref{Distance}(a)--(c) show the CDFs of all distance parameters of static and dynamic scatterers under high, medium, and low VTDs, respectively. From Fig.~\ref{Distance}, it is clear that the CDF of static scatterer distance parameters  matches will with the Gamma distribution. The CDF of the
Gamma distribution is given by
\begin{equation}
F^\mathrm{G}_\mathrm{s}(x)=\frac{\gamma(\alpha^\mathrm{G}_\mathrm{s}, \beta^\mathrm{G}_\mathrm{s} x)}{\Gamma(\alpha^\mathrm{G}_\mathrm{s})}
\end{equation}
where $\alpha^\mathrm{G}_\mathrm{s}$ is the shape parameter and $\beta^\mathrm{G}_\mathrm{s}$ is the rate parameter of the Gamma distribution. $\gamma(\cdot,\cdot)$ is the lower incomplete Gamma  function and $\Gamma(\cdot)$ is the Gamma function.
Different from static scatterers, 
the CDF of dynamic scatterer distance parameters matches will with the Rayleigh distribution. The CDF of the
Rayleigh distribution for the dynamic scatterer  distance parameter is expressed by
\begin{equation}
F^\mathrm{R}_\mathrm{d}(x)=1-e^{-\frac{x^2}{2 (\sigma^\mathrm{R}_\mathrm{d})^2}}
\end{equation}
where $\sigma^\mathrm{R}_\mathrm{d}$ is the scale parameter of the Rayleigh distribution. In Table~\ref{Parameter_1} and Fig.~\ref{Distance},
compared to dynamic scatterers, the distance parameter of static scatterers is larger. This is because that dynamic scatterers are mainly vehicles surrounding the transceiver, while static scatterers are mainly trees and buildings on the roadside, and they are far away from the transceiver. For the impact of VTDs, as the VTD increases, the mean and variance of dynamic scatterer distance parameters increase, whereas the mean and variance of static scatterer distance parameters  increase first and then decrease. The physical reason is that there are more dynamic vehicles around the transceiver in high VTDs, where the propagation environment is more complex and the spatial distribution of dynamic scatterers is more dispersed. Also, the increase in the number of vehicles  leads to a rich SE and more dispersed spatial distribution of static scatterers. Nevertheless, as the number of vehicles further increases, static components via the trees and buildings are greatly blocked by vehicles, leading to more concentrated spatial distribution of static scatterers.

\subsection{Angle Parameters}
\begin{figure*}[!t]
	\centering
	\subfigure[]{\includegraphics[width=0.28\textwidth]{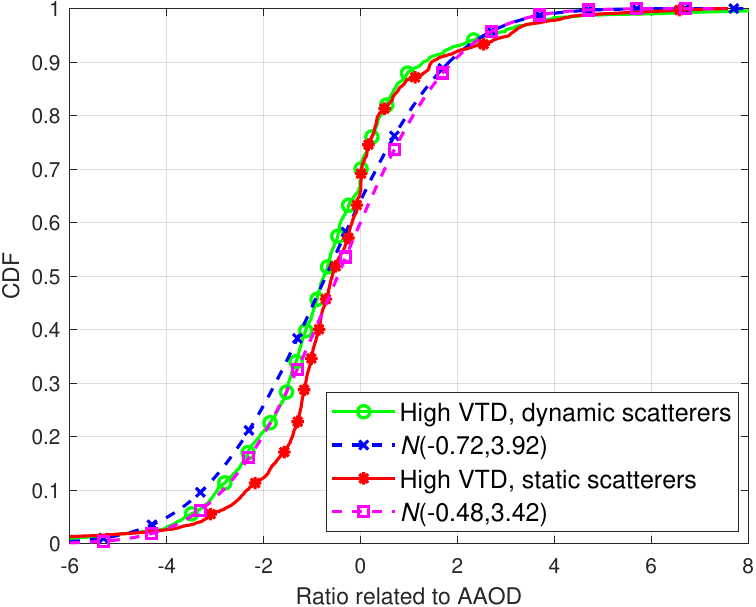}}
	\subfigure[]{\includegraphics[width=0.28\textwidth]{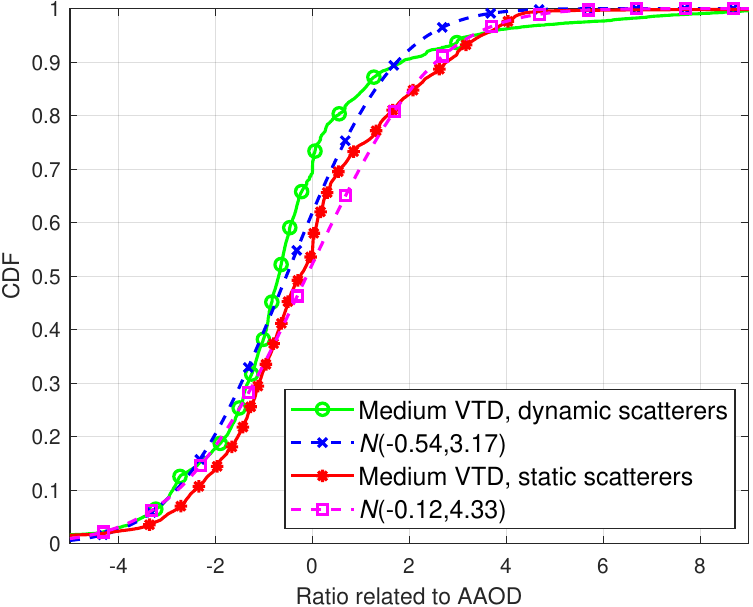}} 
   \subfigure[]{\includegraphics[width=0.28\textwidth]{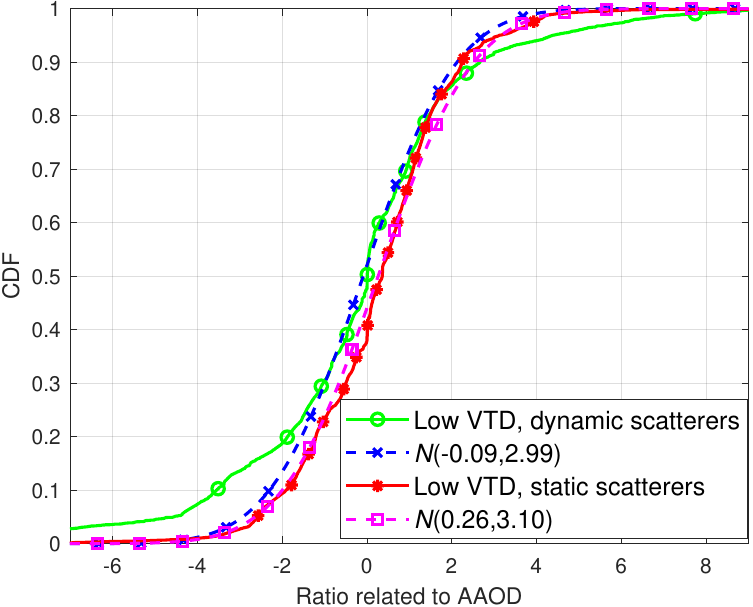}}
	\caption{CDFs of static/dynamic scatterer angle parameters, i.e., AAoD, with the Gaussian distribution fitting under different VTDs. (a) High VTD. (b) Medium VTD. (c) Low VTD.}
 \label{Angle}
\end{figure*}

An in-depth understanding  of angle parameters is also important for channel characterization \cite{IET11}. The authors in \cite{Huang1} found that the angle parameters follow the Laplace
distribution, while the propagation environment was limited to two-dimensional (2D) environment. Furthermore, Yang \emph{et al.} \cite{Yang1} also explored that the angle parameters follow the Laplace distribution in 3D V2V propagation environment, while disregarded the distinction between static  scatterers and dynamic scatterers. Based on the sensing-communication simulation dataset in \cite{dataset_cc}, the angle characteristics, including  azimuth angle of departure (AAoD),  azimuth angle of arrival (AAoA), elevation angle of departure (EAoD), and  elevation angle of arrival (EAoA) of static and dynamic scatterers, are investigated. For clarity, we take the AAoD as an example for analysis. Two new angle parameters ${\alpha}^\mathrm{c1,c3}_{i}(t)$ and ${\alpha}^\mathrm{c1,c3}_{j}(t)$, which represent the ratios of the AAoDs of the $i$-th static scatterer and the $j$-th dynamic scatterer to transceiver distance in the communication link between Car1, i.e., Tx, and Car3, i.e., Rx,  are introduced 
 \begin{equation}
     {\alpha}^\mathrm{c1,c3}_{i}(t)=\frac{\gamma_i^\mathrm{c1,c3}(t)}{\|\mathbf{T}^\mathrm{c1,c3}(t)-\mathbf{R}^\mathrm{c1,c3}(t)\|}
 \end{equation}
  \begin{equation}
     {\alpha}^\mathrm{c1,c3}_{j}(t)=\frac{\gamma_j^\mathrm{c1,c3}(t)}{\|\mathbf{T}^\mathrm{c1,c3}(t)-\mathbf{R}^\mathrm{c1,c3}(t)\|}
 \end{equation}
where $\gamma_i^\mathrm{c1,c3}(t)$ and $\gamma_j^\mathrm{c1,c3}(t)$ denote the AAoDs of the $i$-th static scatterer and the $j$-th dynamic scatterer in the communication link between Car1 and Car3. Similarly, we compute the angle parameter, i.e., AAoD, of each static/dynamic scatterer
for each communication link at each snapshot. Figs.~\ref{Angle}(a)--(c) depict the CDFs of all angle
parameters, i.e., AAoDs, of static and dynamic scatterers under high, medium, and low VTDs, respectively. A close agreement between the Gaussian distribution and the corresponding CDFs is achieved. The CDF of the Gaussian distribution for AAoDs related to static/dynamic scatterers is expressed by
\begin{equation}
    F^\mathrm{AAoD}_\mathrm{s/d}(x)=\frac{1}{2}\left[1+\operatorname{erf}\left(\frac{x-\mu^\mathrm{AAoD}_\mathrm{s/d}}{\sigma^\mathrm{AAoD}_\mathrm{s/d} \sqrt{2}}\right)\right]
\end{equation}
where $\mu^\mathrm{AAoD}_\mathrm{s/d}$ is the mean and $\sigma^\mathrm{AAoD}_\mathrm{s/d}$ is the standard deviation of the 
Gaussian distribution for AAoDs related to static/dynamic scatterers. $\operatorname{erf}(\cdot)$ is the error function. 
Note that other static/dynamic scatterer angle parameters, i.e., AAoA ${\theta}^\mathrm{c1,c3}_{j}(t)$, EAoD ${\beta}^\mathrm{c1,c3}_{j}(t)$, and EAoA ${\phi}^\mathrm{c1,c3}_{j}(t)$, also follow the Gaussian distribution and statistical  values are given in Table~\ref{Parameter_1}. It can be seen from Table~\ref{Parameter_1} and Fig.~\ref{Angle} that, compared with static scatterers, dynamic scatterers have smaller variance in the horizontal angle, while larger variance in the elevation angle. This is because that the spatial distribution of dynamic scatterers is more concentrated and closer to the transceiver. Furthermore, as the VTD increases, the variance of   dynamic scatterer angle parameters increases, whereas the variance of   static scatterer angle parameters   increases first and then decreases. The physical reason is similar to the aforementioned static/dynamic scatterer number and distance parameters.

\subsection{Power-Delay Characteristics}
\begin{figure*}[!t]
	\centering
	\subfigure[]{\includegraphics[width=0.28\textwidth]{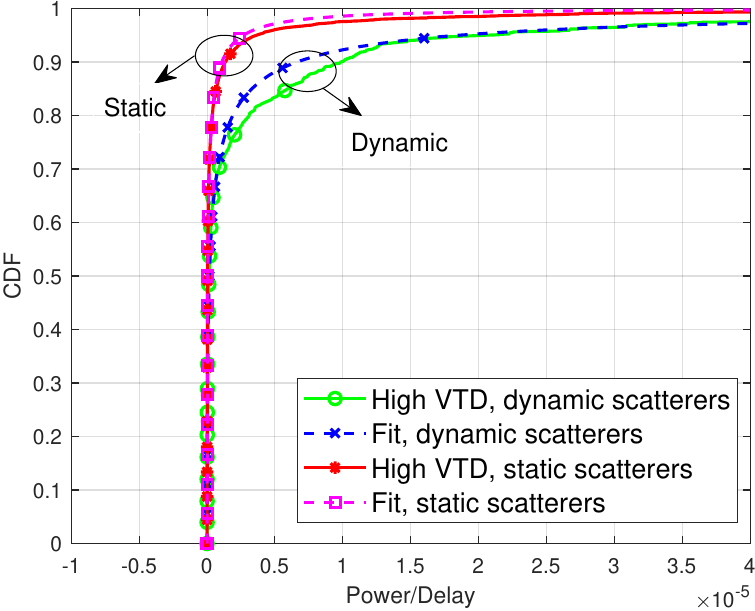}}
	\subfigure[]{\includegraphics[width=0.28\textwidth]{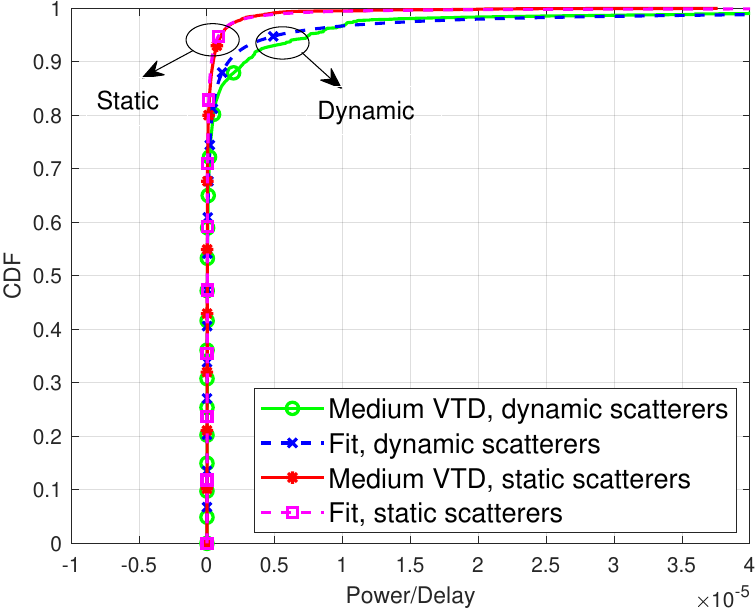}} 
   \subfigure[]{\includegraphics[width=0.28\textwidth]{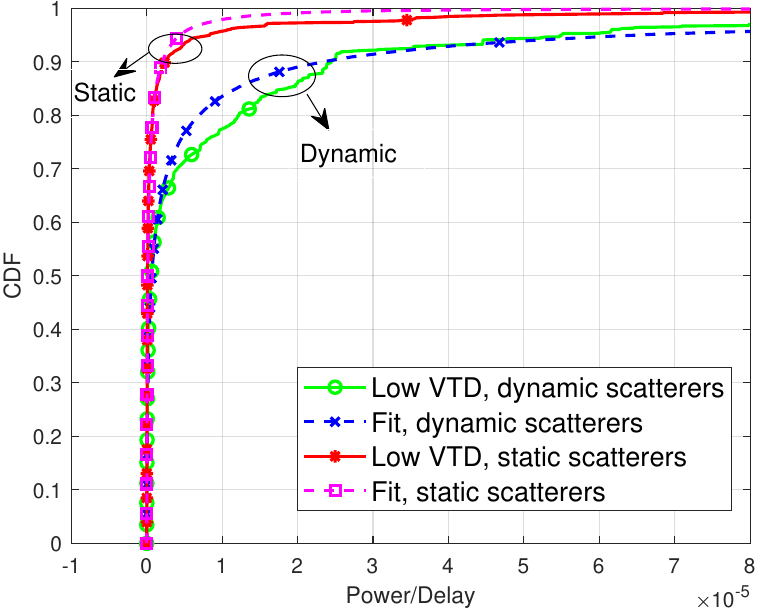}}
	\caption{CDFs of the ratios of static/dynamic scatterer power to static/dynamic scatterer delay  with the Exponential expression fitting under different VTDs. (a) High VTD. (b) Medium VTD. (c) Low VTD.}
 \label{Power}
\end{figure*}

In addition to 3D coordinates of static and dynamic scatterers, the sensing-communication simulation dataset also includes delay and power information of each  path in the propagation environment. 
%Based on the detected static/dynamic scatterers, we define that a path containing only static/dynamic scatterers is a static/dynamic path.
As mentioned in \cite{WINNER}, the path  power is an exponential function of the path delay. Therefore, the power $P_k(t)$ and $P_l(t)$ of the $k$-th static path and the $l$-th dynamic path in the communication link can be expressed by 
\begin{equation}
 P_k(t)=\exp \left(-\xi_\mathrm{s} \tau_k(t)-\eta_\mathrm{s}\right) 10^{-\frac{Z_\mathrm{s}}{10}} 
  \label{Pk}
\end{equation}
\begin{equation}
 P_l(t)=\exp \left(-\xi_\mathrm{d} \tau_l(t)-\eta_\mathrm{d}\right) 10^{-\frac{Z_\mathrm{d}}{10}}  
 \label{Pl}
\end{equation} 
where $\xi_\mathrm{s/d}$ and $\eta_\mathrm{s/d}$ are the introduced delay-related parameters of static/dynamic scatterers. $\tau_{k/l}(t)$ is the delay of the $k/l$-th static/dynamic path. $Z_\mathrm{s}$ and $Z_\mathrm{d}$ obey the Gaussian distributions $\mathcal{N}\left(0, \sigma_\mathrm{E,s}^2\right)$ and $\mathcal{N}\left(0, \sigma_\mathrm{E,d}^2\right)$, respectively. To apply the linear regression based on ordinary least squares, we transform \eqref{Pk} and \eqref{Pl} as
\begin{equation}
 -\mathrm{ln}P_k(t)= \xi_\mathrm{s} \tau_k(t)+\eta_\mathrm{s}+\frac{\ln 10}{10} Z_\mathrm{s} 
\end{equation}
\begin{equation}
 -\mathrm{ln}P_l(t)= \xi_\mathrm{d} \tau_l(t)+\eta_\mathrm{d}+\frac{\ln 10}{10} Z_\mathrm{d}. 
\end{equation}
We calculate the power and delay of each static/dynamic path at each snapshot, and then exploit the linear regression based on ordinary least squares. Table~\ref{Parameter_1} gives the fitted parameters. Figs.~\ref{Power}(a)--(c) further  present the fitting results under high, medium, and low VTDs, respectively, validating the accuracy of the fitted  parameters. Compared to static scatterers, the power of dynamic scatterers is more sensitive to the change of delay, and thus the increase in the delay of dynamic scatterers significantly reduce their power. Furthermore, as the VTD decreases, the decrease in the dynamic scatterer power is more obvious with the increase in the delay, i.e., larger $\xi_\mathrm{d}$.
%This is because that, as VTDs decrease, the number of dynamic scatterers decreases and the delay sensitivity increases. Nonetheless, the decrease in VTDs causes the sensitivity, which is proportional to $\xi_\mathrm{s}$, of the static scatterer power to the delay to decrease first and then increase because the number of static scatterers decreases first and then increases.

\section{A Novel Non-Stationary and Consistent  LA-GBSM}
With the help of LiDAR point clouds, the statistical distributions of
parameters, such as distance, angle, number, and power, related to
static and dynamic scatterers  under high, medium, and low VTD conditions are
obtained. Furthermore, based on the obtained statistical distributions, a novel non-stationary and consistent mmWave LA-GBSM for vehicular intelligent sensing-communication  integration is proposed. It is the first time that non-RF sensory information, which is intelligently processed by machine learning, is exploited to facilitate the RF channel modeling. 
%Since vehicular intelligent sensing-communication  integration is a typical application scenario in SoM, the proposed LA-GBSM can support the SoM research.}

\subsection{Framework of the Proposed LA-GBSM}
\begin{figure}[!t]
		\centering	\includegraphics[width=0.49\textwidth]{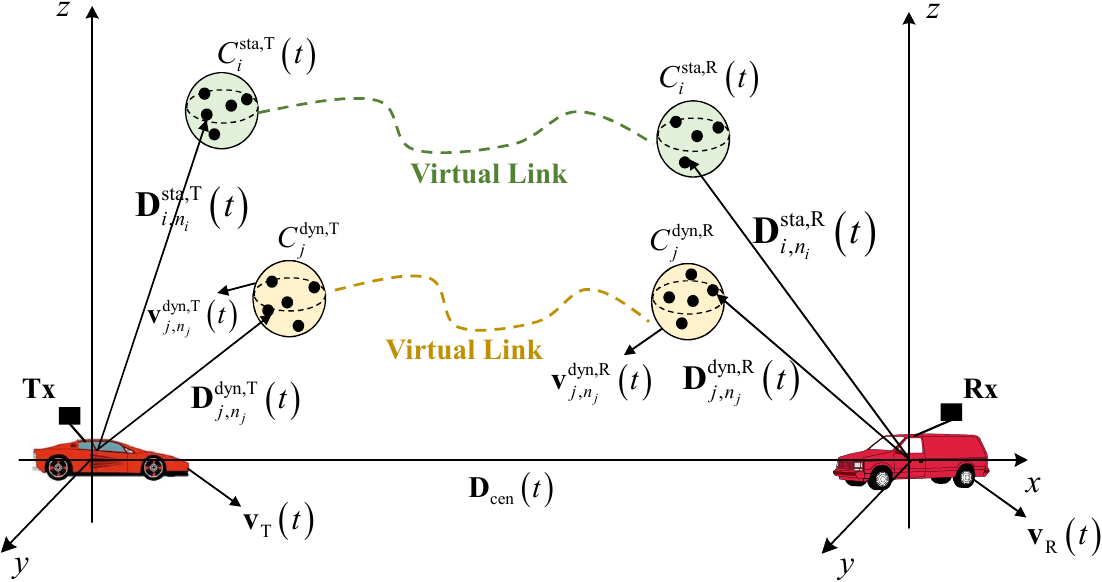}
	\caption{Geometry of the proposed  channel model for vehicular intelligent sensing-communication integration.}
	\label{fig:model}
	\end{figure}

 The geometric representation of the proposed LA-GBSM is depicted in  
Fig.~\ref{fig:model}.  The transceiver distance is $D_\mathrm{cen}(t_0)$ at initial time.
As a result, according to Table~\ref{Parameter_1}, the numbers $N_\mathrm{s}(t_0)$ and $M_\mathrm{s} (t_0)$ of static and dynamic scatterers are  generated by the Logistic distribution. The position of the $i/j$-th static/dynamic scatterer is  generated based on the
  distance $D_i(t_0)$/$D_j(t_0)$ with  the Gamma/Rayleigh distribution and angles, i.e., AAoDs ${\alpha}_{i}(t_0)$ and ${\alpha}_{j}(t_0)$, AAoAs ${\theta}_{i}(t_0)$ and ${\theta}_{j}(t_0)$, EAoDs ${\beta}_{i}(t_0)$ and ${\beta}_{j}(t_0)$, and EAoAs ${\phi}_{i}(t_0)$ and ${\phi}_{j}(t_0)$, with the Gaussian distribution. 
The generated static and dynamic scatterers are clustered via the $K$-Means clustering algorithm to obtain static and dynamic clusters, respectively. Note that the numbers $N_\mathrm{c}(t_0)$ and $M_\mathrm{c}(t_0)$, i.e., $K$ value in the $K$-Means clustering algorithm,  of static and dynamic clusters are also determined by the Logistic distribution  according to  Table~\ref{Parameter_1}. Furthermore, we define the cluster whose centroid  is closer to Tx/Rx as the Tx/Rx cluster. 
The Tx cluster and the Rx cluster are randomly shuffled and matched to form the twin-cluster similar to \cite{Wu}. The transmission
between the twin-cluster is abstracted as a virtual link, where other clusters may exist and introduce the second-order and beyond interaction. 
The $i$-th Tx/Rx static cluster and the $j$-th Tx/Rx dynamic cluster  are represented by $C^\mathrm{sta,T/R}_i$ and $C^\mathrm{dyn,T/R}_j$. 
The dynamic cluster $C^\mathrm{dyn,T/R}_j$ moves with the velocity vector $\mathbf{v}^\mathrm{dyn,T/R}_j(t)$. %The $n_i$-th scatterer within the $i$-th Tx/Rx static cluster is $S^\mathrm{sta,T/R}_{i,n_i}$ and the $n_j$-th scatterer within the $j$-th Tx/Rx dynamic cluster is $S^\mathrm{dyn,T/R}_{i,n_j}$. 
The scatterer within a cluster has the same velocity vector as the cluster, which can be expressed by $\mathbf{v}_{j,n_j}^\mathrm{dyn,T/R}(t)=\mathbf{v}_{j}^\mathrm{dyn,T/R}(t)$.
Finally, the velocity vectors of transceivers are $\mathbf{v}_\mathrm{T}(t)$ and $\mathbf{v}_\mathrm{R}(t)$.

The complex channel gain of the line-of-sight (LoS) component can be represented as
	\begin{equation}
	h^\mathrm{LoS}(t)=Q(t) \mathrm{exp}\left[j 2 \pi  \int_{t_0}^tf^\mathrm{LoS}(t)\mathrm{d}t+j \varphi^\mathrm{LoS}(t) \right]
	\label{LoS}
	\end{equation}
 	where ${T_0}$ denotes the observation time interval and $Q(t)$ represents a rectangular window function \cite{window}, which can be given by
	\begin{align}
	Q(t)=
	\left\lbrace \begin{matrix}
	1,  &  t_0\leqslant t \leqslant T_0,
	\\
	0,  &  \mathrm{otherwise}.
	\end{matrix} \right.
	\end{align}
In addition, the Doppler frequency $f^\mathrm{LoS}(t)$, phase shift $\varphi^\mathrm{LoS}(t)$, as well as delay $\tau^\mathrm{LoS}(t)$ are calculated as $f^\mathrm{LoS}(t)=\frac{1}{\lambda} \frac{\left\langle\mathbf{D}_\mathrm{cen}(t), \mathbf{v}_\mathrm{R}(t)-\mathbf{v}_\mathrm{T}(t)\right\rangle}{\left\|\mathbf{D}_\mathrm{cen}(t)\right\|}$, $\varphi^\mathrm{LoS}(t)=\varphi_{0}+\frac{2 \pi}{\lambda}\left\|\mathbf{D}_\mathrm{cen}(t)\right\|$, and $   \tau^\mathrm{LoS}(t)=\frac{\left\|\mathbf{D}_\mathrm{cen}(t)\right\|}{c}$, where
 $\left\langle \cdot,\cdot \right\rangle$, $\varphi_{0}$, $\lambda$, and $c$ represent the inner product,    initial phase shift, carrier wavelength, and   speed of light. $\mathbf{D}_\mathrm{cen}(t)$ is the transceiver distance at time $t$ and is given as 
\begin{equation}
\mathbf{D}_\mathrm{cen}(t)=\mathbf{D}_\mathrm{cen}(t_0)+\int^t_{t_0}\mathbf{v}_\mathrm{R}(t)\mathrm{d}t-\int^t_{t_0}\mathbf{v}_\mathrm{T}(t)\mathrm{d}t.
\end{equation}
Continuously arbitrary vehicular movement trajectories (VMTs) of transceivers and dynamic clusters are captured similar to our previous work in \cite{TWC_mixed}.

For non-LoS (NLoS) components, the complex channel gain via the static twin-cluster $C^\mathrm{sta,T/R}_{i}$ by the $n_i$-th  static scatterer can be written by
	\begin{equation}
 \begin{scriptsize}
\begin{aligned}
	h^\mathrm{sta}_{i,n_i}(t)=&Q(t) \sqrt{P^\mathrm{sta}_{i,n_i}(t)} \\
 &\times \mathrm{exp}\left\{j 2 \pi \left[ \int_{t_0}^tf^\mathrm{sta,T}_{i,n_i}(t)\mathrm{d}t+\int_{t_0}^tf^\mathrm{sta,R}_{i,n_i}(t)\mathrm{d}t \right]+j \varphi^\mathrm{sta}_{i,n_i}(t)\right\}
	\label{NLoS2}
 \end{aligned}
 \end{scriptsize}
	\end{equation}
where $P^\mathrm{sta}_{i,n_i}(t)$ represents the normalized static scatterer power,  which can be  determined according to Table~\ref{Parameter_1}. The Doppler frequency  of the static cluster at Tx/Rx is computed by $ f^\mathrm{sta,T/R}_{i,n_i}(t)=\frac{1}{\lambda} \frac{\left\langle\mathbf{D}^\mathrm{sta,T/R}_{i,n_i}(t), \mathbf{v}_\mathrm{T/R}(t)\right\rangle}{\left\|\mathbf{D}^\mathrm{sta,T/R}_{i,n_i}(t)\right\|}$,
where $\mathbf{D}^\mathrm{sta,T/R}_{i,n_i}(t)$ denotes the distance between the Tx/Rx and the $n_i$-th static scatterer via the static cluster $C^\mathrm{sta,T/R}_{i}$.
The corresponding delay  is computed by $\tau^\mathrm{sta}_{i,n_i}(t)=\frac{\left\|\mathbf{D}^\mathrm{sta,T}_{i,n_i}(t)\right\|+\left\|\mathbf{D}^\mathrm{sta,R}_{i,n_i}(t)\right\|}{c}+\tilde{\tau}_{i}(t)$,
where $\tilde{\tau}_{i}(t)$ represents the abstracted delay of the virtual link between the static twin-cluster $C^\mathrm{sta,T/R}_{i}$ and follows the Exponential distribution \cite{virtual}. 
The phase shift of the component via the $n_i$-th static scatterer within the static twin-cluster $C^\mathrm{sta,T/R}_{i}$ is given by $\varphi^\mathrm{sta}_{i,n_i}(t)=\varphi_{0}+\frac{2 \pi}{\lambda}\left[\left\|\mathbf{D}_{i,n_i}^\mathrm{sta,T}(t)\right\|+\left\|\mathbf{D}_{i,n_i}^\mathrm{sta,R}(t)\right\|+c\tilde{\tau}_{i}(t)\right]$.
Also, the complex channel gain for NLoS components via the dynamic twin-cluster $C^\mathrm{dyn,T/R}_{j}$ by the $n_j$-th  dynamic scatterer is given as 
\begin{equation}
\begin{scriptsize}
\begin{aligned}
h^\mathrm{dyn}_{j,n_j}(t)=&Q(t) \sqrt{P^\mathrm{dyn}_{j,n_j}(t)} \\
& \times \mathrm{exp}\left\{j 2 \pi \left[ \int_{t_0}^tf^\mathrm{dyn,T}_{j,n_j}(t)\mathrm{d}t+\int_{t_0}^tf^\mathrm{dyn,R}_{j,n_j}(t)\mathrm{d}t \right]+j \varphi^\mathrm{dyn}_{j,n_j}(t)\right\}
 \end{aligned}
 \end{scriptsize}
	\end{equation}
where $P^\mathrm{dyn}_{j,n_j}(t)$ denotes the normalized dynamic scatterer power,  which can be properly determined according to Table~\ref{Parameter_1}. The Doppler frequency  of the dynamic cluster at Tx/Rx is computed by $f^\mathrm{dyn,T/R}_{j,n_j}(t)=\frac{1}{\lambda} \frac{\left\langle\mathbf{D}^\mathrm{dyn,T/R}_{j,n_j}(t), \mathbf{v}_\mathrm{T/R}(t)-\mathbf{v}_{j}^\mathrm{dyn,T/R}(t)\right\rangle}{\left\|\mathbf{D}^\mathrm{dyn,T/R}_{j,n_j}(t)\right\|}$,
where $\mathbf{D}^\mathrm{dyn,T/R}_{j,n_j}(t)$ is the distance between the Tx/Rx and the $n_j$-th dynamic scatterer via the dynamic cluster $C^\mathrm{dyn,T/R}_{j}$.
The corresponding delay  is computed by $\tau^\mathrm{dyn}_{j,n_j}(t)=\frac{\left\|\mathbf{D}^\mathrm{dyn,T}_{j,n_j}(t)\right\|+\left\|\mathbf{D}^\mathrm{dyn,R}_{j,n_j}(t)\right\|}{c}+\tilde{\tau}_{j}(t)$,
where $\tilde{\tau}_{j}(t)$ represents the abstracted delay of the virtual link between the dynamic twin-cluster $C^\mathrm{dyn,T/R}_{j}$ and also follows the Exponential distribution. 
The phase shift of the component via the $n_j$-th dynamic scatterer within the dynamic twin-cluster $C^\mathrm{dyn,T/R}_{j}$ can be expressed as $\varphi^\mathrm{dyn}_{j,n_j}(t)=\varphi_{0}+\frac{2 \pi}{\lambda}\left[\left\|\mathbf{D}_{j,n_j}^\mathrm{dyn,T}(t)\right\|+\left\|\mathbf{D}_{j,n_j}^\mathrm{dyn,R}(t)\right\|+c\tilde{\tau}_{i}(t)\right]$.

The complex channel gain of ground reflection component is written as
\begin{equation}
 \begin{scriptsize}
\begin{aligned}
{h}^\mathrm{GR}(t)
=&Q(t)\sqrt{P^\mathrm{GR}(t)}\\
&\times \mathrm{exp} \left\{j2\pi\left[\int_{t_0}^t{f}^\mathrm{GR,T}(t)\mathrm{d}t+\int_{t_0}^t{f}^\mathrm{GR,R}(t)\mathrm{d}t\right]+j \varphi^\mathrm{GR}(t)\right\}
\label{GRCIR}
\end{aligned}
 \end{scriptsize}
\end{equation}
where  $P^\mathrm{GR}(t)$, ${f}^\mathrm{GR,T/R}(t)$, and $\varphi^\mathrm{GR}(t)$ denote the power, Doppler frequency at Tx/Rx, and phase shift of ground reflection component, respectively.  In addition, $\tau^\mathrm{GR}(t)$ is the delay of ground reflection component.
The power $P^\mathrm{GR}(t)$, Doppler frequency at Tx/Rx ${f}^\mathrm{GR,T/R}(t)$,  phase shift $\varphi^\mathrm{GR}(t)$, and delay $\tau^\mathrm{GR}(t)$  can be computed similar to our previous work in \cite{TWC_mixed}.

Overall, the channel impulse response (CIR) of the proposed LA-GBSM for vehicular intelligent sensing-communication integration can be expressed by \eqref{desde}, shown at the top of the next page. 
\begin{figure*}[!t]
	\begin{equation}
	\begin{footnotesize}
		\begin{aligned}
  h(t,\tau)=&\underbrace{\sqrt{\frac{\Omega(t)}{\Omega(t)+1}}
		h^\mathrm{LoS}(t)\delta\left(\tau-\tau^\mathrm{LoS}(t)\right)}_\mathrm{LoS}+\underbrace{\sqrt{\frac{\eta^\mathrm{GR}(t)}{\Omega(t)+1}}h^\mathrm{GR}(t) \delta\left(\tau-\tau^\mathrm{GR}(t)\right)}_\mathrm{Ground \, Reflection}\\
&+\underbrace{\sum_{i=1}^{N_\mathrm{c}(t)}\sum_{n_i=1}^{N_\mathrm{s}(t)}\sqrt{\frac{\eta^\mathrm{sta}(t)}{\Omega(t)+1}}h^\mathrm{sta}_{i,n_i}(t) \delta\left(\tau-\tau^\mathrm{sta}_{i,n_i}(t)\right)+\sum_{j=1}^{M_\mathrm{c}(t)}\sum_{n_j=1}^{M_\mathrm{s}(t)}\sqrt{\frac{\eta^\mathrm{dyn}(t)}{\Omega(t)+1}}h^\mathrm{dyn}_{j,n_j}(t) \delta\left(\tau-\tau^\mathrm{dyn}_{j,n_j}(t)\right)
	}_\mathrm{NLoS}.
\label{desde}
		\end{aligned}
		\end{footnotesize}
	\end{equation}
 		\hrulefill
\vspace*{4pt}
\end{figure*}
 $\Omega(t)$ represents the Ricean factor. $\eta^\mathrm{GR}(t)$,
$\eta^\mathrm{sta}(t)$, and 
$\eta^\mathrm{dyn}(t)$ represent the power ratios of the ground reflection component, component via static clusters, and  component via dynamic clusters and have $\eta^\mathrm{GR}(t)+\eta^\mathrm{sta}(t)+\eta^\mathrm{dyn}(t)=1$.  In the derived CIR, the key channel-related parameters, including number, distance, angle, and power, under high, medium, and low VTD conditions are   determined  according to Table~\ref{Parameter_1}.

\subsection{Capturing of Channel Non-Stationarity and Consistency}
Channel non-stationarity is a typical channel characteristic  and channel consistency is an inherent channel physical feature. As mentioned in \cite{myCOMST}, channel non-stationarity and  channel consistency need to be captured in high-mobility  V2V communication channels under the mmWave frequency band.  Based on
the obtained statistical distributions, a new VR-based algorithm in consideration of newly generated static/dynamic clusters is developed to capture channel  non-stationarity and consistency in the time domain, i.e., time non-stationarity and consistency. Moreover, a frequency-dependent factor is introduced to model channel non-stationarity in the frequency domain, i.e., frequency non-stationarity.
%Therefore, time-frequency non-stationarity of channels with time consistency are mimicked in the proposed channel model for sensing and communication integration.

Currently, the VR method is widely used in channel modeling \cite{VR00}.
In the VR method, clusters are visible and contribute to  channel realization only if they are within the VR. As the VR and clusters move, the visible cluster set smoothly changes, and thus time non-stationarity and consistency are captured.
To mimic smooth static/dynamic  cluster time evolution in  channels for vehicular intelligent sensing-communication integration, a new VR-based algorithm under high, medium, and low VTDs is developed based on the obtained statistical distribution of distance parameters. The VR is set to the 3D ellipsoid with the transceiver as the focus.  For the VR, i.e., 3D ellipsoid, assigned to the static/dynamic cluster, the minor axis $2b^\mathrm{sta/dyn}(t)$ and the focal length $2c^\mathrm{sta/dyn}(t)$ are equal to the distance $D_\mathrm{cen}(t)$ between  transceivers, which can be calculated by the positions of  transceivers. Meanwhile, the major axis  $2a^\mathrm{sta/dyn}(t)$ is equal to the sum of the distances from the static/dynamic cluster to  transceivers and is computed 
 based on the obtained statistical distribution of distance parameters, as listed in Table~\ref{Parameter_1}. Then,   
 a  visibility factor $\epsilon^\mathrm{sta/dyn}$   is introduced to determine the  major axis  $2a^\mathrm{sta/dyn}(t)$ of VR assigned to static/dynamic clusters. For  dynamic clusters, based on the CDF of the Rayleigh distribution of their distance parameters, the major axis  $2a^\mathrm{dyn}(t)$ of VR is computed by 
\begin{equation}
    1-\exp \left(-\frac{\left(\frac{2a^\mathrm{dyn}(t)-D_\mathrm{cen}(t)}{D_\mathrm{cen}(t)}\right)^2}{2 {\sigma_\mathrm{d}^{\mathrm{R}}}^2}\right)=\epsilon^\mathrm{dyn}.
\end{equation}
Then, the major axis $2a^\mathrm{dyn}(t)$ of VR can be computed by 
\begin{equation}
 2a^\mathrm{dyn}(t)=\sqrt{-2 {\sigma_\mathrm{d}^{\mathrm{R}}}^2 \ln (1-\epsilon^\mathrm{dyn})} D_\mathrm{cen}(t)+D_\mathrm{cen}(t).
\end{equation}
Unlike dynamic clusters, for static clusters, based on the CDF of the Gamma distribution of their  distance parameters, the major axis  $2a^\mathrm{sta}(t)$ of VR can be computed by
\begin{equation}
    \frac{1}{\Gamma(\alpha^\mathrm{G}_\mathrm{s})} \gamma\left(\alpha^\mathrm{G}_\mathrm{s}, \beta^\mathrm{G}_\mathrm{s}\left(\frac{2a^\mathrm{sta}(t)-D_\mathrm{cen}(t)}{D_\mathrm{cen}(t)}\right)\right)=\epsilon^\mathrm{sta}.
\end{equation}
Since there is no closed-form solution for the major axis  $2a^\mathrm{sta}(t)$, we calculate the numerical  solution. Therefore, the VRs assigned to static and dynamic clusters are the 3D ellipsoids determined by the precise RT-based communication data, and  static and dynamic clusters have different and time-varying VRs.
Since the statistical values $\sigma_\mathrm{d}^{\mathrm{R}}$, $\alpha^\mathrm{G}_\mathrm{s}$, and $\beta^\mathrm{G}_\mathrm{s}$, as listed in Table~\ref{Parameter_1}, include  high, medium, and low VTDs, the VRs assigned to static and dynamic clusters also consider high, medium, and low VTDs. With the help of VRs assigned to static and dynamic clusters, the numbers $N^\mathrm{v}_\mathrm{c}(t)$ and $M^\mathrm{v}_\mathrm{c}(t)$ of visible static and dynamic clusters can be determined. 

Based on the derived statistical distribution of number parameters,  the developed algorithm further models newly generated static/dynamic clusters. We take the static cluster as an example.
For a certain distance $D_\mathrm{cen}(t)$ between  Tx and Rx at time $t$, the number parameter  $N^\mathrm{L}_\mathrm{c}(t)$ related to static clusters is randomly generated via the Logistic distribution with the statistical values as listed in  Table~\ref{Parameter_1}. 
If the value $N^\mathrm{L}_\mathrm{c}(t)$  is greater than the number $N^\mathrm{v}_\mathrm{c}(t)$ of visible static clusters, the number of newly generated static clusters is given by
\begin{equation}
N^\mathrm{new}_\mathrm{c}(t)=N^\mathrm{L}_\mathrm{c}(t)-N^\mathrm{v}_\mathrm{c}(t).    
\end{equation}
 In this case, there are $N_\mathrm{c}(t)=N^\mathrm{L}_\mathrm{c}(t)$ static clusters that contribute to channel realization.
On the contrary, if the value $N^\mathrm{L}_\mathrm{c}(t)$  is less than the number $N^\mathrm{v}_\mathrm{c}(t)$ of visible static clusters,
the number of newly generated static clusters is $N^\mathrm{new}_\mathrm{c}(t)=0$ and there are $N_\mathrm{c}(t)=N^\mathrm{v}_\mathrm{c}(t)$  static clusters that contribute to channel realization. Note that the determination of newly generated dynamic clusters has the same steps as that of newly generated static clusters.
Consequently, the numbers of newly generated static and dynamic clusters are  determined via the  statistical distribution of cluster number parameters. Then, the parameters of newly generated static and dynamic clusters are randomly generated according to Table~\ref{Parameter_1}.
In Algorithm 1, the pseudo code of the developed algorithm 
is given.

\begin{algorithm}[t!]
	\caption{A Novel VR-Based Algorithm with Newly Generated Static and Dynamic Clusters}
\begin{footnotesize}
	\begin{algorithmic}[1]
  	\While {$t<T_\mathrm{end}$}
  		\State Determine 3D ellipsoid  for static cluster via Gamma distribution;
     \For{each static cluster $i\in [1,N_\mathrm{c}(t)]$} 
     	\If {Static cluster is within  3D ellipsoid}
      \State Visible static cluster that contributes to CIR;
      \Else
      \State Invisible static cluster  that cannot contribute to CIR;
      	\EndIf
		\EndFor
   		\State Determine 3D ellipsoid  for dynamic cluster via Rayleigh distribution;
     \For{each dynamic cluster $j\in [1,M_\mathrm{c}(t)]$} 
     	\If {Dynamic cluster is within  3D ellipsoid}
      \State Visible dynamic cluster that contributes to CIR;
      \Else
      \State Invisible dynamic cluster  that cannot contribute to CIR;
      	\EndIf
		\EndFor
     \State Randomly generate number parameter $N^\mathrm{L}_\mathrm{c}(t)$ for static cluster via  Logistic distribution;
        \If {$N^\mathrm{L}_\mathrm{c}(t)$ $>$ $N^\mathrm{v}_\mathrm{c}(t)$ }
              \State Generated static cluster number $N^\mathrm{new}_\mathrm{c}(t)=N^\mathrm{L}_\mathrm{c}(t)-N^\mathrm{v}_\mathrm{c}(t)$;
                    \Else
      \State  Generated static cluster number $N^\mathrm{new}_\mathrm{c}(t)=0$;
      	\EndIf
            \State Randomly generate number parameter $M^\mathrm{L}_\mathrm{c}(t)$ for dynamic cluster via  Logistic distribution;
        \If {$M^\mathrm{L}_\mathrm{c}(t)$ $>$  $M^\mathrm{v}_\mathrm{c}(t)$ }
              \State Generated dynamic cluster number $M^\mathrm{new}_\mathrm{c}(t)=M^\mathrm{L}_\mathrm{c}(t)-M^\mathrm{v}_\mathrm{c}(t)$;
                    \Else
      \State Generated dynamic cluster number  $M^\mathrm{new}_\mathrm{c}(t)=0$;
      	\EndIf
        \State Randomly generate parameters of  generated static and dynamic clusters according to Table~\ref{Parameter_1};
    \State $t=t+\Delta t$;	
		\EndWhile
	\end{algorithmic}
	\end{footnotesize}
\end{algorithm}

%Currently, the VR method is widely used to mimic channel non-stationarity and consistency. In \cite{Flordelis}, the VR was set to the  circle based on  channel measurement data, while the VR was 2D and ignored the impact of VTD. To overcome this limitation, our previous work in \cite{TWC_mixed} 

The use of mmWave technology results in the necessity of capturing  frequency non-stationarity \cite{myCOMST}. To this aim, a frequency-dependent factor is introduced and used to the  time-varying transfer function (TVTF). The TVTF is  derived by using the Fourier transform to CIR $h(t, \tau)$ in respect of delay $\tau$, which is written by
\begin{equation}
 H'(t, f)=\int_{-\infty}^{\infty} h(t, \tau) \mathrm{exp}\left({-j 2 \pi f \tau}\right) \mathrm{d} \tau.   
\end{equation}
To capture the frequency-dependent path gain, the introduced frequency-dependent factor is applied to the TVTF, which is expressed as \eqref{eee}, shown at the top of the next page.
\begin{figure*}[!t]
	\begin{equation}
	\begin{scriptsize}
 	\begin{aligned}
	H(t, f)&= \underbrace{\sqrt{\frac{\Omega(t)}{\Omega(t)+1}} h^{\mathrm{LoS}}(t) \mathrm{exp}\left[{-j 2 \pi f \tau^{\mathrm{LoS}}(t)}\right]}_\mathrm{LoS} +\underbrace{\sqrt{\frac{\eta^\mathrm{GR}(t)}{\Omega(t)+1}} \left(\frac{f}{f_{c}}\right)^{\chi} h^\mathrm{GR}(t) \mathrm{exp}\left[{-j 2 \pi f\tau^\mathrm{GR}(t)}\right]}_\mathrm{Ground \, Reflection}\\
&~~+\underbrace{\sqrt{\frac{\eta^\mathrm{sta}(t)}{\Omega(t)+1}} \left(\frac{f}{f_{c}}\right)^{\chi}  \sum_{i=1}^{N_\mathrm{c}(t)}\sum_{n_i=1}^{N_\mathrm{s}(t)} h^\mathrm{sta}_{i,n_i}(t) \mathrm{exp}\left[{-j 2 \pi f\tau^\mathrm{sta}_{i,n_i}(t)}\right]
 +\sqrt{\frac{\eta^\mathrm{dyn}(t)}{\Omega(t)+1}} \left(\frac{f}{f_{c}}\right)^{\chi} \sum_{j=1}^{M_\mathrm{c}(t)}\sum_{n_j=1}^{M_\mathrm{s}(t)} h^\mathrm{dyn}_{j,n_j}(t) \mathrm{exp}\left[{-j 2 \pi f\tau^\mathrm{dyn}_{j,n_j}(t)}\right]}_\mathrm{NLoS}
 \label{eee}
		\end{aligned}
		\end{scriptsize}
	\end{equation}
 		\hrulefill
\vspace*{4pt}
\end{figure*}
 $\chi$ is a frequency-dependent parameter that is dependable  on  the environment \cite{ewf2w}. With the aid of the frequency-dependent factor, the frequency-dependent path gain is imitated, thus modeling frequency non-stationarity.

\section{Channel Statistical Properties}

Key statistical properties of channels for vehicular intelligent sensing-communication integration are derived.

\subsection{Time-Frequency Correlation Function}
Based on the TVTF, the TF-CF is computed by 
\begin{equation}
\xi(t,f;\Delta{t},\Delta{f})=\mathbb{E}[H^\ast(t,f)H(t+\Delta{t},f+\Delta{f})]
\label{eq:CF}
\end{equation}
where $\mathbb{E}[\cdot]$ and $(\cdot)^*$ represent the expectation operation and complex conjugate operation, respectively. Moreover, the TF-CFs of LoS components,  ground reflection components, and NLoS components via static and dynamic clusters are given by \eqref{eq:CF_sum} with \eqref{eq:zLoS}--\eqref{eq:zNLoS}, shown at the top of the next page.
\begin{figure*}[!t]
\begin{equation}
\begin{aligned}
\xi(t,f;\Delta{t},\Delta{f})=\xi^\mathrm{LoS}(t,f;\Delta{t},\Delta{f})+\xi^\mathrm{GR}(t,f;\Delta{t},\Delta{f})+\xi^\mathrm{NLoS}(t,f;\Delta{t},\Delta{f})
\end{aligned}
	\label{eq:CF_sum}
	\end{equation}
 \begin{equation}
\begin{scriptsize}
\label{eq:zLoS}
\begin{aligned}
	\xi^{\mathrm{LoS}}\left(t, f ; \Delta t, \Delta f\right)=\sqrt{\frac{\Omega(t)\Omega(t+\Delta t)}{(\Omega(t)+1)(\Omega(t+\Delta t)+1)}}h^{\mathrm{LoS} *}(t) h^{\mathrm{LoS}}(t+\Delta t) \exp  \left[j 2 \pi f \tau^{\mathrm{LoS}}(t)-(f+\Delta f) \tau^{\mathrm{LoS}}(t+\Delta t)\right]
	\end{aligned}
	\end{scriptsize}
	\end{equation}
  \begin{equation}
\begin{scriptsize}
\label{eq:zLoS2}
\begin{aligned}
	\xi^{\mathrm{GR}}\left(t, f ; \Delta t, \Delta f\right)=\sqrt{\frac{\eta^\mathrm{GR}(t)\eta^\mathrm{GR}(t+\Delta t)}{(\Omega(t)+1)(\Omega(t+\Delta t)+1)}}h^{\mathrm{GR} *}(t) h^{\mathrm{GR}}(t+\Delta t) \exp  \left[j 2 \pi f \tau^{\mathrm{GR}}(t)-(f+\Delta f) \tau^{\mathrm{GR}}(t+\Delta t)\right]
	\end{aligned}
	\end{scriptsize}
	\end{equation}
 	\begin{equation}
	\begin{scriptsize}
	\begin{aligned}
&\xi^\mathrm{NLoS}(t,f;\Delta{t},\Delta{f})\\
&~=\sqrt{\frac{\eta^\mathrm{sta}(t)\eta^\mathrm{sta}(t+\Delta t)}{(\Omega(t)+1)(\Omega(t+\Delta t)+1)}}{\mathbb{E}\left[\sum_{i=1}^{N_\mathrm{c}(t)}\sum_{i'=1}^{N_\mathrm{c}(t+\Delta{t})}\sum_{n_i=1}^{N_\mathrm{s}(t)}\sum_{n'_s=1}^{N_\mathrm{s}(t+\Delta{t})} h_{i,n_i}^\mathrm{sta\ast}(t)h_{i',n'_i}^\mathrm{sta}(t+\Delta{t})\mathrm{exp}\left({j2\pi \tau^\mathrm{sta}_{i,n_i}(t)f-(f+\Delta{f})\tau^\mathrm{sta}_{i',n'_i}(t+\Delta{t})}\right)\right]}\\
&~+\sqrt{\frac{\eta^\mathrm{dyn}(t)\eta^\mathrm{dyn}(t+\Delta t)}{(\Omega(t)+1)(\Omega(t+\Delta t)+1)}}{\mathbb{E}\left[\sum_{j=1}^{M_\mathrm{c}(t)}\sum_{j'=1}^{M_\mathrm{c}(t+\Delta{t})}\sum_{n_j=1}^{M_\mathrm{s}(t)}\sum_{n'_j=1}^{M_\mathrm{s}(t+\Delta{t})} h_{j,n_j}^\mathrm{dyn\ast}(t)h_{j',n'_j}^\mathrm{dyn}(t+\Delta{t})\mathrm{exp}\left(j2\pi \tau^\mathrm{dyn}_{j,n_j}(t)f-(f+\Delta{f})\tau^\mathrm{dyn}_{j',n'_j}(t+\Delta{t})\right) \right]}.
	\end{aligned}
	\end{scriptsize}
	\label{eq:zNLoS}
	\end{equation}
 		\hrulefill
\vspace*{4pt}
\end{figure*}
Based on the TF-CF, the time auto-correlation function (TACF) and the frequency correlation function (FCF) can be derived by setting $\Delta{f}=0$ and $\Delta{t}=0$, respectively.

\subsection{Doppler Power Spectral Density}
By taking the Fourier transfer of the derived TACF, the DPSD can be expressed by
	\begin{equation}
	\begin{aligned}
	\Theta(t;f_\mathrm{D})=\int_{-\infty}^{+\infty}\xi(t;\Delta{t})e^{-j2\pi{f_\mathrm{D}}\Delta{t}}\mathrm{d}(\Delta{t})
	\end{aligned}
	\end{equation}
where $\xi(t;\Delta{t})$ is the TACF  and $f_\mathrm{D}$ is the Doppler frequency. The derived time-varying DPSD indicates the time-varying characteristic of the proposed LA-GBSM for vehicular intelligent sensing-communication integration.

\section{Simulation Results and Analysis}
Key channel statistical properties are simulated and compared with the accurate RT-based results.  The carrier frequency is $f_\mathrm{c}=28$~GHz with $2$~GHz communication bandwidth.  The numbers of antennas at Tx and Rx are $M_\mathrm{T}=M_\mathrm{R}=1$.
The azimuth and elevation angles at Tx and Rx sides are  $\phi_\mathrm{T}^\mathrm{E}=\phi_\mathrm{R}^\mathrm{E}=\pi/4$, $\theta_\mathrm{T}^\mathrm{A}=\pi/3$, and $\theta_\mathrm{R}^\mathrm{A}=3\pi/4$.  Delays of virtual links $\tilde{\tau}_{i}(t)$ and $\tilde{\tau}_{j}(t)$ obey the Exponential distribution with the mean and variance $80$~ns and $15$~ns. The environment-dependent factor is $\chi=1.35$ \cite{ewf2w}. Unless otherwise stated, the aforementioned  parameter remains unchanged.

\subsection{Model Simulation}
\begin{figure}[!t]
		\centering	\includegraphics[width=0.38\textwidth]{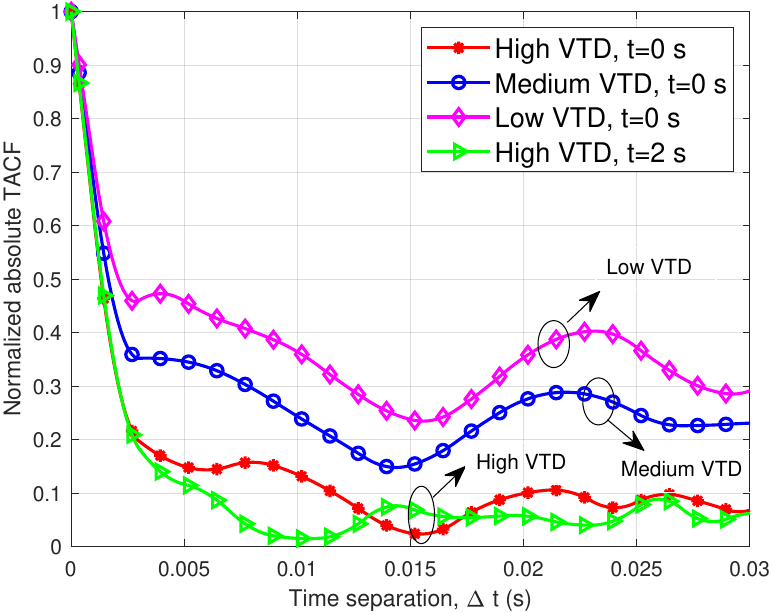}
	\caption{TACFs under different VTDs and time instants ($D_\mathrm{cen}(t_0)=80$~m,  $v_\mathrm{T}(t_0)=18$~m/s,  $v_\mathrm{R}(t_0)=15$~m/s, $\bar{v}^\mathrm{dyn,T}_{j}(t_0)=14$~m/s, $\bar{v}^\mathrm{dyn,R}_{j}(t_0)=15$~m/s).}
	\label{TACF_VTD}
	\end{figure}

Fig.~\ref{TACF_VTD} shows the absolute normalized TACFs under high, medium, and low VTDs at $t=0$~s and  $t=2$~s. In Fig.~\ref{TACF_VTD}, TACFs depend on the time instant as well as the time separation, and hence time non-stationarity is characterized. Moreover, the TACF  significantly decreases as the VTD increases. This is because that, as the number of mobile vehicles increases, the  channels are more complex and dynamic, and the temporal correlation is lower.

 \begin{figure}[!t]
		\centering	\includegraphics[width=0.38\textwidth]{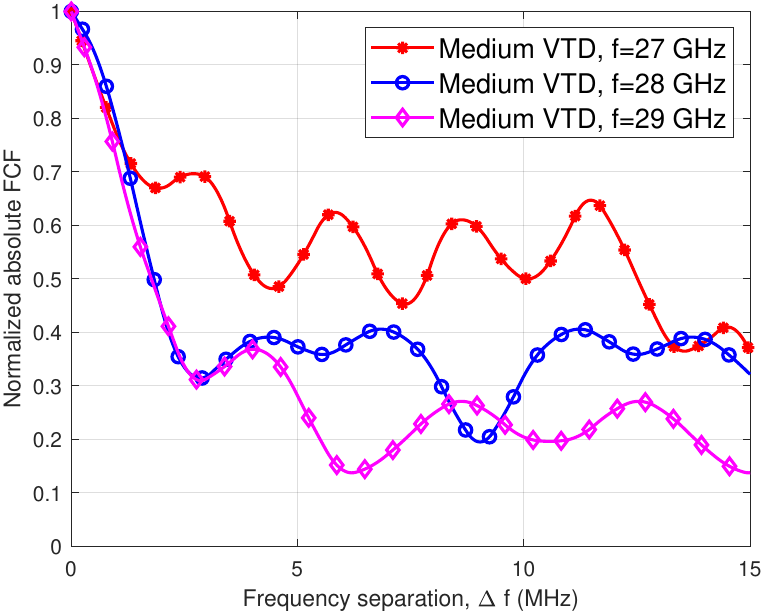}
	\caption{FCFs under different frequencies ($D_\mathrm{cen}(t_0)=50$~m,  $v_\mathrm{T}(t_0)=15$~m/s,  $v_\mathrm{R}(t_0)=17$~m/s, $\bar{v}^\mathrm{dyn,T}_{j}(t_0)=\bar{v}^\mathrm{dyn,R}_{j}(t_0)=8$~m/s, $t=0$~s).}
	\label{FCF}
	\end{figure}
 
Fig.~\ref{FCF} depicts absolute normalized FCFs under  frequencies $f=27$~GHz,  $f=28$~GHz, and $f=29$~GHz at the medium VTD. It can be observed from Fig.~\ref{FCF} that FCFs are dependable on the frequency and the frequency separation,  demonstrating frequency non-stationarity of the proposed LA-GBSM. Furthermore, compared to $27$~GHz, the FCF under higher frequency is lower. This phenomenon can be explained that  channels at higher frequency are more complex and thus have lower frequency correlation.

\subsection{Model Verification}
\begin{figure}[!t]
		\centering	\includegraphics[width=0.38\textwidth]{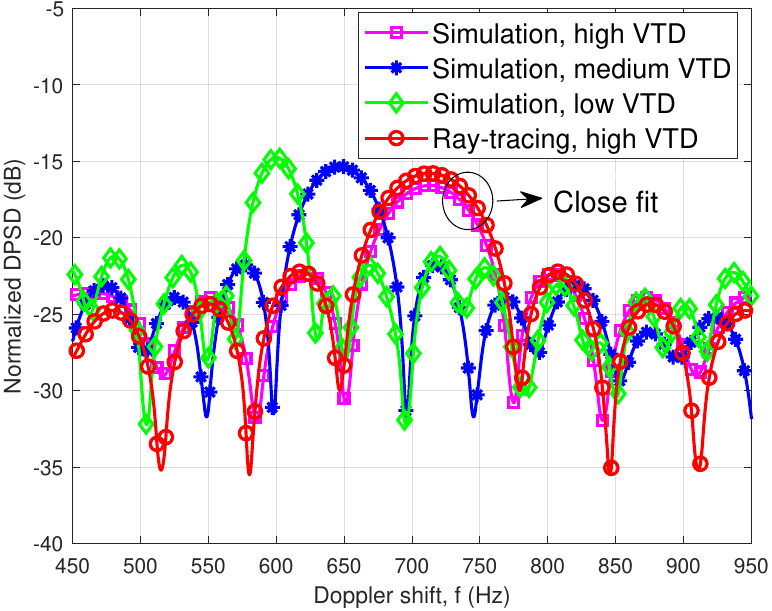}
	\caption{Comparison of simulated DPSDs and RT-based DPSDs ($D_\mathrm{cen}(t_0)=83.6$~m,  $v_\mathrm{T}(t_0)=10$~m/s,  $v_\mathrm{R}(t_0)=6$~m/s, $\bar{v}^\mathrm{dyn,T}_{j}(t_0)=\bar{v}^\mathrm{dyn,R}_{j}(t_0)=8$~m/s).}
	\label{DPSD_fit}
	\end{figure}
Aiming at verifying the generality of the proposed LA-GBSM, we obtain the  RT-based CIR  collected in Wireless InSite \cite{WI} and the scenario is shown in Fig.~\ref{WI_AirSim}. Note that the method of comparing  simulation results of the proposed channel model with  RT-based results to verify the generality of the proposed channel model is widely used.

The RT-based CIR under the high VTD is processed to obtain the RT-based DPSD, which is compared with the simulated DPSDs under high, medium, and low VTDs  in Fig.~\ref{DPSD_fit}. The close match between the RT-based DPSD and the simulated DPSD under the high VTD is achieved. Moreover, compared to low VTDs, the distributions of DPSDs under high and medium VTDs are flatter. This phenomenon is explained that  the received power in the high VTD
tends to come from mobile vehicles around the transceiver over all directions.
 
\begin{figure}[!t]
		\centering	\includegraphics[width=0.38\textwidth]{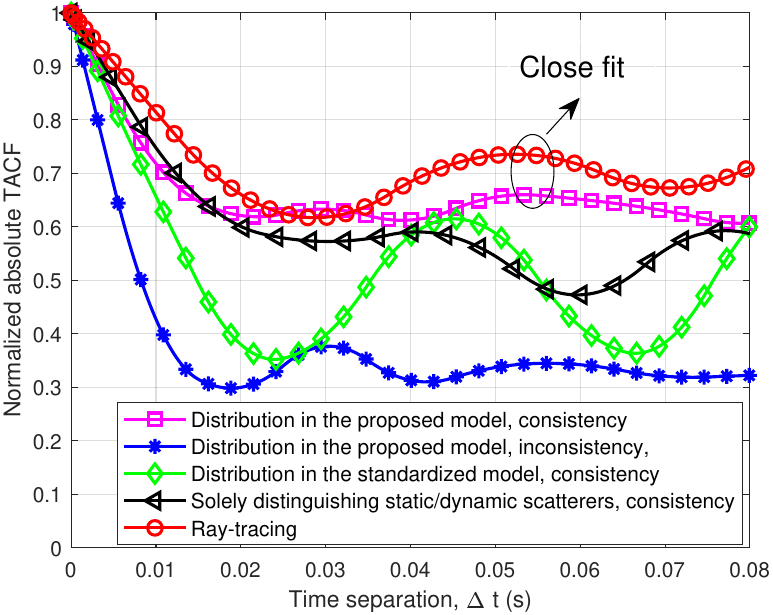}
	\caption{Comparison of simulated TACFs and RT-based TACFs ($D_\mathrm{cen}(t_0)=54.8$~m,  $v_\mathrm{T}(t_0)=8.33$~m/s,  $v_\mathrm{R}(t_0)=10$~m/s, $\bar{v}^\mathrm{dyn,T}_{j}(t_0)=\bar{v}^\mathrm{dyn,R}_{j}(t_0)=12$~m/s).}
	\label{TACF_fit}
	\end{figure}

In low VTDs, the RT-based CIR is further processed to obtain the RT-based TACF. In Fig.~\ref{TACF_fit}, the RT-based TACF is compared with the  TACF of the model solely distinguishing  static/dynamic scatterers, the TACF of the model with the parameter obeying the distribution derived in the standardized model, e.g., \cite{3GPP}, and the TACF of the model under time inconsistency. To imitate  time inconsistency, the birth-death process method is used to mimic the appearance and disappearance of clusters. It can be seen that, only the simulated TACF with the parameter obeying the distribution derived in the proposed  model under   time consistency can fit the RT-based TACF well.  For the simulated TACF under time inconsistency, the smooth channel evolution in the time domain is ignored and the  channel correlation at  adjacent moments is underestimated, hence exhibiting lower TACF. For the model solely distinguishing  static/dynamic scatterers and the model with the parameter obeying the distribution derived in the standardized model, their TACFs cannot fit well with the RT-based TACF as statistical distributions of parameters related to static/dynamic scatterers are not modeled accurately.

\section{Conclusions}
This paper has developed a novel channel modeling approach, i.e., LA-GBSM. Based on the developed LA-GBSM approach, a new non-stationary and consistent mmWave channel model has been   proposed to support the  design of vehicular intelligent sensing-communication integration systems in ITSs. The proposed LA-GBSM has been  parameterized under different VTD conditions by a sensing-communication simulation dataset. The distance parameters of static and dynamic scatterers have followed the Gamma and Rayleigh distributions, respectively. Although the number, angle, and power-delay characteristics of static and dynamic scatterers have respectively obeyed the same Logistic, Gaussian, and Exponential distributions, their statistical distribution values have shown a significant difference.
Furthermore,  based on the obtained statistical distribution under high, medium, and low VTD conditions, a new VR-based algorithm in consideration of newly generated static/dynamic clusters has been developed to capture
 channel non-stationarity and  consistency. Key channel statistics have been derived and simulated. Simulation results have shown that the time-frequency non-stationarity and time consistency have been captured. Compared to the medium and low VTD conditions, the proposed LA-GBSM for vehicular intelligent sensing-communication integration has exhibited lower temporal correlation and flatter distribution of DPSD under the high VTD condition. By comparing simulation results and RT-based results, the generality of the proposed LA-GBSM has been verified.

\ifCLASSOPTIONcaptionsoff
  \newpage
\fi

\end{document}